\documentclass[runningheads,envcountsame]{llncs}

\usepackage{ifluatex}
\ifluatex
  \usepackage{fontspec}
  \defaultfontfeatures{Ligatures=TeX}
  \usepackage[disable,positioning=sLeaderNorthEastBelowStacks,leadertype=sBezier,splitting=weightedMedian]{luatodonotes}
\else
  \usepackage[utf8]{inputenc}
  \usepackage[T1]{fontenc}
  \usepackage[disable]{todonotes}
\fi

\usepackage[english]{babel}
\usepackage[font=small,labelfont=bf,labelsep=period]{caption}
\usepackage[font=small,labelfont=normalfont]{subcaption}

\usepackage{amsmath,amssymb,amsfonts}
\usepackage{graphicx}
\usepackage{enumerate}
\usepackage{xspace}
\usepackage{xparse}
\usepackage{multicol}
\usepackage{etoolbox}
\usepackage{xstring}
\usepackage{booktabs}
\usepackage[binary-units=true]{siunitx}
\usepackage{textcomp}
\usepackage[figuresleft]{rotating}

\makeatletter
\@ifundefined{textsubscript}{
\DeclareRobustCommand*\textsubscript[1]{%
  \@textsubscript{\selectfont#1}}
\def\@textsubscript#1{%
  {\m@th\ensuremath{_{\mbox{\fontsize\sf@size\z@#1}}}}}%
}{}
\makeatother

\usepackage{bold-extra}
\newcommand{\cpp}{{C\nolinebreak[4]\hspace{-.05em}\raisebox{.4ex}{\tiny\bf ++}}}
\newcommand{\python}{\mbox{Python}}
\newcommand{\minisat}{\texttt{mini\-sat}}
\newcommand{\programname}[1]{\textsc{#1}}
\newcommand{\aDP}{\programname{Fpt}}
\newcommand{\aID}{\programname{ItD}}
\newcommand{\aSAT}{\programname{Sat}}

\newcommand{\numchars}{\ensuremath{\kappa}}

\usepackage[bookmarksopen,bookmarksdepth=3]{hyperref}
\newcommand\rurl[1]{\href{http://#1}{\nolinkurl{#1}}}

\usepackage{tikz}
\usetikzlibrary{decorations.pathmorphing,decorations.pathreplacing,decorations.shapes,fit,positioning,shapes.geometric,calc,intersections,plotmarks}
\usetikzlibrary{external}
\tikzsetfigurename{paper-figure}
\tikzset{external/export=false} %
\tikzset{external/verbose IO=true}
\tikzset{external/verbose optimize=true}

\usepackage[]{arxiv-appendix}
\arxivid{XXX}
\arxivcite{arxiv-version}
\arxivauxname{paper-arxiv}
\lncsfalse

\title{Computing Storyline Visualizations \\ with Few Block Crossings%
\iflncs%
  \thanks{\arxivrefthanks}
\else%
  \thanks{Appears in the Proceedings of the 25th International Symposium on Graph Drawing and Network Visualization (GD 2017).}
\fi%
}

\author{Thomas~C.~van~Dijk \and
  Fabian~Lipp\thanks{F.~Lipp was supported by Cusanuswerk. ORCID:~\href{http://orcid.org/0000-0001-7833-0454}{\tt orcid.org/0000-0001-7833-0454}}
  \and Peter~Markfelder
  \and Alexander~Wolff\thanks{ORCID:~\href{http://orcid.org/0000-0001-5872-718X}{\tt orcid.org/0000-0001-5872-718X}}}

\authorrunning{T.~C.~van~Dijk et al.}
\titlerunning{Computing Storyline Visualizations with Few Block Crossings}

\institute{%
  Lehrstuhl f\"ur Informatik I, Universit\"at W\"urzburg, Germany\\
  \rurl{www1.informatik.uni-wuerzburg.de/en/staff},
  \email{first.last@uni-wuerzburg.de}}

\definecolor{cb-Dark2-1}{RGB}{27,158,119}
\definecolor{cb-Dark2-2}{RGB}{217,95,2}
\definecolor{cb-Dark2-3}{RGB}{117,112,179}
\definecolor{cb-Dark2-4}{RGB}{231,41,138}
\definecolor{cb-Dark2-5}{RGB}{102,166,30}
\definecolor{cb-Dark2-6}{RGB}{230,171,2}
\definecolor{cb-Dark2-7}{RGB}{166,118,29}
\definecolor{cb-Dark2-8}{RGB}{102,102,102}
\definecolor{cb-Dark2-9}{RGB}{41,201,201}
\definecolor{cb-Dark2-10}{RGB}{41,201,84}
\definecolor{cb-Dark2-11}{RGB}{201,190,41}
\definecolor{cb-Dark2-12}{RGB}{201,116,41}
\definecolor{cb-Dark2-13}{RGB}{226,25,99}
\definecolor{cb-Dark2-14}{RGB}{25,132,226}

\tikzset{
  meeting/.append style={ultra thick},
  char/.append style={very thick},
  char1/.append style={char,cb-Dark2-1},
  char2/.append style={char,cb-Dark2-2},
  char3/.append style={char,cb-Dark2-3},
  char4/.append style={char,cb-Dark2-4},
  char5/.append style={char,cb-Dark2-5},
  char6/.append style={char,cb-Dark2-6},
  char7/.append style={char,cb-Dark2-7},
  char8/.append style={char,cb-Dark2-8},
  char9/.append style={char,cb-Dark2-9},
  char10/.append style={char,cb-Dark2-10},
  char11/.append style={char,cb-Dark2-11},
  char12/.append style={char,cb-Dark2-12},
  char13/.append style={char,cb-Dark2-13},
  char14/.append style={char,cb-Dark2-14},
}

\newcounter{x}
\makeatletter
\pgfmathdeclarefunction{mydim}{1}{%
  \begingroup
    \pgfmath@count=0\relax
    \expandafter\pgfmath@mydim@i\pgfutil@firstofone#1\pgfmath@token@stop
    \edef\pgfmathresult{\the\pgfmath@count}%
    \pgfmath@smuggleone\pgfmathresult%
  \endgroup}
\def\pgfmath@mydim@i#1{%
    \ifx\pgfmath@token@stop#1%
    \else
      \advance\pgfmath@count by 1\relax
      \expandafter\pgfmath@mydim@i
    \fi}

\providecommand*\@nameundef[1]{%
  \expandafter\let\csname #1\endcsname\@undefined}
\pgfkeys{
  /tikz/execute style/.style = {#1},
  /tikz/execute macro/.style = {execute style/.expand once=#1}
}
\pgfkeys{
  /storyline/.is family, /storyline,
  drawingstyle/.estore in = \storyline@drawingstyle,
  meetingstyle/.estore in = \storyline@meetingstyle,
  meetingopacity/.estore in = \storyline@meetingopacity,
  leftwidth/.estore in = \storyline@leftwidth,
  rightwidth/.estore in = \storyline@rightwidth,
  crossinglength/.estore in = \storyline@crossinglength,
  distx/.estore in = \storyline@distx,
  xgrid/.estore in = \storyline@xgrid,
  disty/.estore in = \storyline@disty,
  label/.estore in = \storyline@label,
  meetingaddy/.estore in = \storyline@meetingaddy,
  meetingaddx/.estore in = \storyline@meetingaddx,
  labelshifty/.estore in = \storyline@labelshifty,
  name path/.initial,
  l1/.style={drawingstyle=char1,label=$1$},
  l2/.style={drawingstyle=char2,label=$2$},
  l3/.style={drawingstyle=char3,label=$3$},
  l4/.style={drawingstyle=char4,label=$4$},
  l5/.style={drawingstyle=char5,label=$5$},
  l6/.style={drawingstyle=char6,label=$6$},
  l7/.style={drawingstyle=char7,label=$7$},
  l8/.style={drawingstyle=char8,label=$8$},
  l9/.style={drawingstyle=char9,label=$9$},
  l10/.style={drawingstyle=char10,label=$10$},
  l11/.style={drawingstyle=char11,label=$11$},
  l12/.style={drawingstyle=char12,label=$12$},
  l13/.style={drawingstyle=char13,label=$13$},
  l14/.style={drawingstyle=char14,label=$14$},
  invisible/.code={\def\storyline@invisible{true}},
  norightlabel/.code={\def\storyline@norightlabel{true}},
  moveinstantaneousright/.code={\def\storyline@moveinstantaneousright{true}},
  default/.style = {
    drawingstyle={},
    meetingstyle={fill=black,draw=none},
    meetingopacity={0.2},
    leftwidth={0},
    rightwidth={0},
    crossinglength={0.5}, %
    distx={6},
    xgrid={},
    disty={0.5},
    label={},
    meetingaddy={0.3},
    meetingaddx={1},
    labelshifty={}
  }
}
\newcommand*{\storylineset}{\pgfqkeys{/storyline}}
\storylineset{default}
\newcounter{storyline@tmp}
\newcommand{\storylineaccumulategrid}[1][0]{
  \setcounter{storyline@tmp}{0}
  \xdef\storyline@xgrid@accu{}
  \xdef\curx{0}
  \foreach \xstep in \storyline@xgrid {
    \pgfmathparse{\xstep+\curx}
    \xdef\curx{\pgfmathresult}
    \xappto\storyline@xgrid@accu{\curx,}
    \stepcounter{storyline@tmp}
  }
  \loop
    \ifnum\value{storyline@tmp}<#1
    \pgfmathparse{\storyline@distx+\curx}
    \xdef\curx{\pgfmathresult}
    \xappto\storyline@xgrid@accu{\curx,}
    \stepcounter{storyline@tmp}
  \repeat
  \typeout{accu: \storyline@xgrid@accu} %
}
\newcommand{\drawstoryline}[2][]{
  \begin{scope}[/storyline, #1]
  \setcounter{x}{0}
  \@nameundef{lasty}
  \@nameundef{firsty}
  \@nameundef{firstx}
  \xdef\lastx{0}
  \xdef\curx{0}
  \xdef\pathtodraw{}
  \pgfmathtruncatemacro{\storyline@xgrid@length}{int(mydim(\storyline@xgrid))}
  \typeout{length: \storyline@xgrid@length} %
  \foreach \y [count=\c,
    evaluate=\c as \xstep using {\ifnum\c>\storyline@xgrid@length{\storyline@distx}\else{\storyline@xgrid}[\c-1]\fi}
    ]
    in {#2}
  {
    \xdef\lastx{\curx}
    \pgfmathparse{\xstep+\curx}
    \xdef\curx{\pgfmathresult}

    \IfStrEq{\y}{x}{
      \stepcounter{x}
    }{
      \@ifundefined{firstx}{
        \xdef\firstx{\curx}
        \xdef\firstxstep{\xstep}
      }{}
      \@ifundefined{lasty}{
        \xdef\lasty{\y}
        \xdef\firsty{\y}
        \IfStrEq{\storyline@leftwidth}{x}{}{
          \xappto\pathtodraw{({\curx-\xstep*\storyline@leftwidth},
              \storyline@disty*\lasty)
            -- (\curx,\storyline@disty*\lasty)}
        }
      }{
        \IfStrEq{\pathtodraw}{}{
          \xappto\pathtodraw{(\lastx,\storyline@disty*\lasty)}
        }{}
        \xappto\pathtodraw{-- ({\lastx+((1-\storyline@crossinglength)*\xstep/2)},
              \storyline@disty*\lasty)
          .. controls ({\lastx+\xstep/2},
              \storyline@disty*\lasty)
          and ({\lastx+\xstep/2},
              \storyline@disty*\y)
          .. ({\curx-((1-\storyline@crossinglength)*\xstep/2)},
              \storyline@disty*\y)
          -- (\curx,\storyline@disty*\y)}
        \xdef\lasty{\y}
        \stepcounter{x}
      }
    }
  }
  \IfStrEq{\storyline@rightwidth}{x}{}{ %
    \xappto\pathtodraw{-- +({\storyline@rightwidth*\storyline@distx},0)}
  }
  \IfStrEq{\pgfkeysvalueof{/storyline/name path}}{}{
    \@ifundefined{storyline@invisible}{
      \draw[\storyline@drawingstyle] \pathtodraw;
    }{
      \path[] \pathtodraw;
    }
  }{
    \@ifundefined{storyline@invisible}{
      \draw[\storyline@drawingstyle,
          name path global={\pgfkeysvalueof{/storyline/name path}}]
        \pathtodraw;
    }{
      \path[name path global={\pgfkeysvalueof{/storyline/name path}}] \pathtodraw;
    }
  }

  \@ifundefined{storyline@invisible}{
    \xdef\storyline@shiftstr{}
    \IfStrEq{\storyline@labelshifty}{}{}{
        \xdef\storyline@shiftstr{+ (0,\storyline@labelshifty)}
    }
    \IfStrEq{\storyline@leftwidth}{x}{}{
      \xdef\storyline@pos{($ ({\firstx-\firstxstep*\storyline@leftwidth},
          \storyline@disty*\firsty) - (.3333em,0) \storyline@shiftstr $)}
      \node[\storyline@drawingstyle,anchor=east,inner sep=0]
        at \storyline@pos
        {\storyline@label};
    }
    \IfStrEq{\storyline@rightwidth}{x}{}{
      \@ifundefined{storyline@norightlabel}{
        \xdef\storyline@pos{($ ({\storyline@distx*
            (\thex-1+\storyline@crossinglength+\storyline@rightwidth)},
            \storyline@disty*\lasty) + (.3333em, 0) \storyline@shiftstr $)}
        \node[\storyline@drawingstyle,anchor=west,inner sep=0]
          at \storyline@pos
          {\storyline@label};
      }{}
    }
  }{}
  \end{scope}
}
\newcommand{\drawstorylinemeeting}[4][]{
  \begin{scope}[/storyline, #1]
  \draw[meeting,#1] ({\storyline@xgrid@accu}[#2],{\storyline@disty*(#3-0.3)})
    -- ({\storyline@xgrid@accu}[#2],{\storyline@disty*(#4+0.3)});
  \end{scope}
}
\newcommand{\drawstorylineintervalmeeting}[3][]{
  \begin{scope}[/storyline, #1]
  \@nameundef{lasty}
  \@nameundef{firsty}
  \@nameundef{firstx}
  \xdef\lastx{0}
  \xdef\curx{0}
  \xdef\storyline@lower{}
  \xdef\storyline@upper{}
  \pgfmathtruncatemacro{\storyline@xgrid@length}{int(mydim(\storyline@xgrid))}
  \foreach \y [count=\c,
    evaluate=\c as \xstep using {\ifnum\c>\storyline@xgrid@length{\storyline@distx}\else{\storyline@xgrid}[\c-1]\fi}
    ]
    in {#3}
  {
    \xdef\lastx{\curx}
    \pgfmathparse{\xstep+\curx}
    \xdef\curx{\pgfmathresult}

    \IfStrEq{\y}{x}{
    }{
      \@ifundefined{firstx}{
        \xdef\firstx{\curx}
        \xdef\firstxstep{\xstep}
      }{}
      \@ifundefined{lasty}{
        \xdef\lasty{\y}
        \xdef\firsty{\y}
      }{
        \IfStrEq{\storyline@lower}{}{
          \xappto\storyline@lower{(\lastx+\storyline@meetingaddx,
              \storyline@disty*\lasty-\storyline@meetingaddy)}
          \xappto\storyline@upper{(\lastx+\storyline@meetingaddx,
              {\storyline@disty*(\lasty+#2-1)+\storyline@meetingaddy})}
        }{}

        \def\lastylower{\storyline@disty*\lasty-\storyline@meetingaddy}
        \def\ylower{\storyline@disty*\y-\storyline@meetingaddy}
        \def\lastyupper{{\storyline@disty*(\lasty+#2-1)+\storyline@meetingaddy}}
        \def\yupper{{\storyline@disty*(\y+#2-1)+\storyline@meetingaddy}}
        \def\cplower{\lastx+\xstep/3}
        \def\cpupper{\lastx+2*\xstep/3}
        \ifnum\y>\lasty
          \def\cplower{\lastx+2*\xstep/3}
          \def\cpupper{\lastx+\xstep/3}
        \fi
        \xappto\storyline@lower{
          -- ({\lastx+((1-\storyline@crossinglength)*\xstep/2)},
              \lastylower)
          .. controls ({\cplower},
              \lastylower)
          and ({\cplower},
              \ylower)
          .. ({\curx-((1-\storyline@crossinglength)*\xstep/2)},
              \ylower)
          -- (\curx-\storyline@meetingaddx,
              \ylower)}
        \xpreto\storyline@upper{
          (\curx-\storyline@meetingaddx,\yupper)
          -- ({\curx-((1-\storyline@crossinglength)*\xstep/2)},
              \yupper)
          .. controls ({\cpupper},
              \yupper)
          and ({\cpupper},
              \lastyupper)
          .. ({\lastx+((1-\storyline@crossinglength)*\xstep/2)},
              \lastyupper) -- }
        \xdef\lasty{\y}
      }
    }
  }
  \IfStrEq{\storyline@lower}{}{
    \begin{scope}[opacity=\storyline@meetingopacity,transparency group]
      \@ifundefined{storyline@moveinstantaneousright}{
        \fill[execute macro=\storyline@meetingstyle]
          (\firstx-1.5*\storyline@meetingaddx,{\storyline@disty*\firsty-\storyline@meetingaddy})
          -- ++({3*\storyline@meetingaddx},0)
          -- ++(0,{\storyline@disty*(#2-1)+2*\storyline@meetingaddy})
          -- ++({-3*\storyline@meetingaddx},0)
          -- cycle;
        }{
          \fill[execute macro=\storyline@meetingstyle]
            (\firstx+\storyline@meetingaddx,{\storyline@disty*\firsty-\storyline@meetingaddy})
            -- ++({2*\storyline@meetingaddx},0)
            -- ++(0,{\storyline@disty*(#2-1)+2*\storyline@meetingaddy})
            -- ++({-2*\storyline@meetingaddx},0)
            -- cycle;
        }
    \end{scope}
  }{
    \begin{scope}[opacity=\storyline@meetingopacity,transparency group]
      \fill[execute macro=\storyline@meetingstyle] \storyline@lower -- \storyline@upper -- cycle;
    \end{scope}
  }
  \end{scope}
}
\newcommand{\drawstorylinegrid}{
  \makeatletter
  \foreach \x [count=\c] in \storyline@xgrid@accu {
    \draw[gray] (\x,-5) -- ++(0,100);
    \node[gray,anchor=east,rotate=90] at (\x,-5) {\tiny\c};
  }
  \makeatother
}
\tikzset{use path/.code=\tikz@addmode{\pgfsyssoftpath@setcurrentpath#1}}
\makeatother

\storylineset{
    norightlabel,
    leftwidth=0.1,
    rightwidth=0.25,
    distx=4,
    crossinglength=0.75,
    meetingaddx=0.75,
    disty=10,
    meetingaddy=3.25,
}

\newcommand{\CS}{\ensuremath{\mathcal{S}}\xspace}
\newcommand{\CE}{\ensuremath{\mathcal{E}}\xspace}

\newcommand{\CM}{\ensuremath{\mathcal{M}}\xspace}
\newcommand{\CL}{\ensuremath{\mathcal{L}}\xspace}
\newcommand{\R}{\ensuremath{\mathbb{R}}\xspace}
\newcommand{\IR}{\ensuremath{\mathbb{I}\mathbb{R}}\xspace}

\begin{document}

\maketitle

\begin{abstract}
  Storyline visualizations show the structure of a story, by
	depicting the interactions of the characters over time.
	Each character is represented by
  an $x$-monotone curve from left to right, and a meeting is
  represented by having the curves of the participating characters
  run close together for some time. There have been various approaches
	to drawing storyline visualizations in an automated way. In order
	to keep the visual
  complexity low, rather than minimizing pairwise crossings of
  curves, we count \emph{block crossings}, that is, pairs
  of intersecting bundles of lines.

  Partly inspired by the ILP-based approach of Gronemann et al.\ [GD 2016]
	for minimizing the number of pairwise crossings, we model the problem
  as a satisfiability problem (since the straightforward ILP
  formulation becomes more complicated and harder to solve).
	Having restricted ourselves to a decision problem, we can apply
	powerful SAT solvers to find optimal drawings in reasonable time.
	We compare this SAT-based approach with two	exact algorithms
	for block crossing minimization, using both the benchmark
	instances of Gronemann et al.\ and random instances. We show that the
	SAT approach is suitable	for real-world instances and identify cases where
	the other algorithms are preferable.
\end{abstract}

\section{Introduction}
\label{sec:introduction}

A storyline visualization is a particular abstraction of the structure of a
narrative.
A good visualization reveals the underlying structure
by removing the details of how the story is presented
and, instead, focusing on which entities %
interact as time passes within the narrative.
This type of diagram was originally conceived to visualize meetings between characters in movies and, though it has since been interpreted more generally as an elegant way to visualize a sequence of interconnected interactions over time, the term storyline visualization remains.

In a storyline visualization, each character
is represented by an $x$-monotone curve in the plane;
we will refer to curves and characters interchangeably.
Time goes from left to right, and a meeting
between a set of characters (occurring for the duration of a given time interval) is represented by a corresponding region in the plane where those
curves come closely together.
This drawing style is commonly attributed
to Munroe~\cite{m-mnc-09}, who represented several popular movies in this fashion.
See Figure~\ref{fig:thematrix} for an example drawn using our system.

\paragraph{Block Crossings in Storyline Visualization.}
When formalizing the drawing of storyline visualizations as an optimization
problem, it is natural to minimize the number of crossings among
the characters.
As with graph drawing in general, this is not the be-all-end-all objective.
For example, two groups of curves crossing each other in a grid structure 
are easier to understand visually than the same number of crossings scattered wildly throughout
the drawing.
In this paper we continue the study of such \emph{block crossings} in storyline visualization.

Intuitively, a block crossing consists of two sets of locally parallel
curves intersecting each other without any further curves in the crossing
area~\cite{dfflmrsw-bcsv-GD16}.
(A formal definition is given below.)
In the design of his movie narrative charts, Munroe seems
aware (at least implicitly) of the concept of block crossings.
Indeed, the Gestalt principle of ``continuity'' or ``good
continuation''~\cite{wertheimer1923untersuchungen} suggests that
block crossings are easier to read, but what exactly makes the most
readable drawing should be analyzed in proper user studies. Here
we focus on practical computational aspects, having decided to
minimize block crossings.

\paragraph{Concurrent Meetings.}
An important modeling decision that the literature has handled variously is whether it is possible for multiple meetings to occur at overlapping time intervals.
Some papers define the input to the storyline visualization problem such that these \emph{concurrent} meetings are impossible, for example by representing the meetings as a totally ordered set.
Whether or not it is important to support concurrent meetings is open for discussion.
One could, for example, represent each scene
of a movie as a separate meeting that includes precisely the characters that participate: then meetings do not overlap.
However, this is a rather mechanical interpretation of what storyline
visualizations are for.  Indeed, rather than strictly following the
order of appearance in the movies, Munroe's ``movie
narrative charts''~\cite{m-mnc-09} visualize the spatio-temporal
structure underlying the story, rather than the presentation of the
story: the $x$-axis in his charts represents time within the story, not
time in the movie.\footnote{For example, Gandalf meets \'Eomer
  \emph{while} the host of elves arrives at Helm's Deep.  In the movie,
  we learn about this only afterward; Munroe's
  visualization of The Lord of the Rings~\cite{m-mnc-09} makes
  clear that this is concurrent.}  This paper
supports concurrent meetings. %

\paragraph{Previous Work.}
Tanahashi and Ma~\cite{tm-dcosv-TVCG12} computed
storyline visualizations automatically and 
discuss various aesthetic criteria to be optimized.
Kim et al.~\cite{kch-tgdt-AVI10} used
storylines to visualize genealogical data: meetings correspond to
marriages and special techniques are used to indicate child--parent
relationships.  

Kostitsyna et al.~\cite{knpss-omcisv-GD15} formalized the problem of
crossing minimization for storylines.  Their aim was to
minimize the number of pairwise crossings (that is, \emph{not} block crossings)
in storylines.
They proved the problem NP-hard, presented an FPT algorithm, and
gave an upper bound on the number of crossings in a restricted setting.
Gronemann et al.~\cite{Gronemann2016} designed an integer linear
program (ILP) to minimize the number of pairwise crossings and
evaluated it experimentally.
Their approach is able to solve instances with 10--20 characters and
up to about 50 meetings from real-world movies (and to a lesser
degree, books) to optimality in a few seconds.

In an earlier paper~\cite{dfflmrsw-bcsv-GD16}, we introduced the concept
of minimizing block crossings for drawing storylines.  We showed that
block crossing minimization in storylines is NP-hard.
For special cases, we provided an approximation algorithm.
Of particular relevance to the current paper are two exact
algorithms, one of which is fixed-parameter tractable (FPT) in the number
of characters.

The current paper improves on the above in two ways.
Firstly, we have developed a new SAT-based algorithm for computing optimal storyline visualizations.
We note that SAT formulations have been used before in graph drawing, for example by Bekos et al.~\cite{bkz-bepss-GD15}.
Whereas we previously restricted ourselves to meetings that are points in time, we now handle concurrent meetings.
(This more general problem is clearly NP-hard as well.)
Secondly, we have now implemented the exact algorithms of our earlier paper~\cite{dfflmrsw-bcsv-GD16}, which enables an experimental evaluation and comparison.
We see that the new algorithm is able to handle larger realistic instances than our previous algorithms, but that the FPT algorithm also has practical relevance.

\paragraph{Problem Definition.}
We generalize the problem statement compared to our previous
paper~\cite{dfflmrsw-bcsv-GD16} in order to handle the instances
used by Gronemann et al.~\cite{Gronemann2016}.
In this more general statement, we support meetings that span a certain
amount of time (instead of allowing only instantaneous meetings); thus,
meetings can overlap with other meetings.
Additionally, we allow for birth and death of characters, that is, each
character is only drawn in the storyline during its \emph{lifespans}
(being a set of time intervals).

A storyline $\CS$ is a triple $(C, M, E)$ where $C = \{1, \dots, \numchars\}$
is a set of \emph{characters},
$M = \{m_1, m_2, \dots, m_n\}$ is a set of meetings, and $E \colon
C \rightarrow \mathcal{P}(\IR)$ describes the lifespans of a
character with $\IR$ being the set of intervals of real numbers.
A \emph{meeting} $m_j$ is a triple $(s_j, e_j, C_j)$ where $s_j \in \R$ is the
start time of the meeting, $e_j \in \R$ is the end time of the meeting, $s_j <
e_j$, and $C_j \subseteq C$ contains the involved characters.
A meeting $m_j$ is said to be \emph{active} at time $t \in \R$ if $t \in [s_j,
e_j)$.
The set $E(i) = \{[b_i^1, d_i^1), \dots, [b_i^{\eta_i}, d_i^{\eta_i})\}$
contains the lifespans of character $i$, that is,
$\eta_i$ disjoint time intervals, in which the character is alive.
For each of these time intervals ($1 \leq \iota \leq \eta_i$), $b_i^\iota$
describes the ``birth'' while $d_i^\iota$
describes the ``death'' of the character.
Character $i$ is said to be \emph{alive} at time $t \in \R$ if $t \in
I$ for some $I \in E(i)$.

We forbid that a character participates in two meetings at the same
time: in our drawing style, it wouldn't be possible to distinguish
the two groups.  More formally, for any two meetings $m_j, m_\ell$
with $s_j < s_\ell < e_j$, we require that $C_j \cap C_\ell =
\emptyset$.  Obviously, character $i$ can only be part of a meeting
$m_j$ if $i$ is alive during the time span of the meeting, that is, if
$[s_j,e_j)\subseteq I$ for some $I \in E(i)$.
In particular, a character cannot be born or die during a meeting.

A solution for a storyline instance $\CS = (C, M, E)$ consists of a
sequence $\Pi = [\pi_1, \dots, \pi_\lambda]$ of permutations of subsets of $C$
and a nondecreasing function $A \colon \R \rightarrow \{1, \dots, \lambda\}$
describing the connection between points in time and the permutations in the
solution.  A solution is \emph{admissible} if it fulfills the
following conditions.
\begin{enumerate}[(a)]
\item For any point in time $t \in \R$, 
  \begin{enumerate}[(i)]
  \item $\pi_{A(t)}$ contains exactly the characters that are alive at
    time~$t$, and
  \item for any meeting that is active at time~$t$, its set of
    characters must be a contiguous block in~$\pi_{A(t)}$.
  \end{enumerate}
\item
  For $p \in \{2,\dots,\lambda\}$:
  \begin{enumerate}[(i)]
  \item If the character sets of~$\pi_{p-1}$ and~$\pi_p$ are
    identical, then either $\pi_{p-1}$ and~$\pi_p$ are identical or
    they differ in a \emph{block crossing}, that is, two adjacent
    blocks of characters switch their order.  Suppose that, after
    renumbering, $\pi_{p-1} = \langle 1, \dots, a, \dots, b, \dots, c,
    \dots, \numchars \rangle$.  Then exchanging the two adjacent blocks
    $\langle a,\dots,b \rangle$ and $\langle b+1,\dots,c \rangle$
    yields the permutation $\pi_p = \langle 1, \dots, a-1, b+1, \dots,
    c, a, \dots, b, c+1, \dots, \numchars \rangle$.
  \item If the character sets of~$\pi_{p-1}$ and~$\pi_p$ are not
    identical, then their intersection must be in the same order
    in~$\pi_{p-1}$ and in~$\pi_p$.
    They need not remain contiguous.
  \end{enumerate}
\end{enumerate}
Now we can formally state the problem that we consider in this paper,
\emph{General Storyline Block Crossing Minimization}: Given a
storyline instance $(C, M, E)$, find an admissible solution $(\Pi, A)$
that minimizes the number of block crossings.

We define $\CE$ to be the finite set of \emph{events}, that is, points
in time, at which a meeting starts or ends or a character is born or dies.
Our aim is to find the smallest number $\lambda_\mathrm{OPT}$ of permutations that
accommodates all events subject to the constraints above. %
This also minimizes the number of block
crossings $\mathrm{bc}_\mathrm{OPT}$ since $\mathrm{bc}_\mathrm{OPT} = \lambda_\mathrm{OPT} - |\CE'| + 1$, where $\CE'$
denotes the points in time at which at least one character is born or dies
(including the birth of the first character and death of the last character in
the storyline).

\paragraph{Our Results.}
Partly inspired by the ILP from Gronemann et al.~\cite{Gronemann2016}, we developed a
SAT formulation of the problem that can be used to decide whether there is a
solution using a fixed number of permutations (and, hence, block
crossings); see Section~\ref{sec:sat}.
Initial experiments with a similar ILP model performed poorly and led us to
explore SAT solvers.
We experimentally compare our new SAT approach to the
two exact algorithms from our previous paper~\cite{dfflmrsw-bcsv-GD16}; see
Section~\ref{sec:experiments}.
The source code of all three implementations is available online\footnote{\url{http://www1.pub.informatik.uni-wuerzburg.de/pub/data/storylines/}}.

\section{SAT Formulation for the Decision Problem}
\label{sec:sat}

We present a SAT formulation that encodes, for a given storyline~$\CS$
and an integer~$\lambda$, whether there is a solution whose sequence of
permutations consists of exactly $\lambda$ elements.
From a satisfying truth assignment we can derive the solution for $\CS$.
The optimal number of block crossings can then be found using this
decision problem by searching for the minimum satisfiable $\lambda$,
for example using linear or exponential search.
  Our formulation is inspired by the ILP of Gronemann et
al.~\cite{Gronemann2016}, which minimizes the number of pairwise crossings in a
storyline visualization.

In the following, we do not always describe the clauses in conjunctive normal form,
using other operators where this improves readability.
The transformation into conjunctive normal form is straightforward.
For the sake of completeness, the result of this transformation is shown in
Appendix~\ref{app:sat-complete}.
In the following, unless specified or bound otherwise, the variables and clauses are
quantified over
all $i,j,k \in C$ with $i \neq j$, $i \neq k$, $j \neq k$ and all $r,p \in
\{1, \dots, \lambda\}$, $r \neq p$, and $\ell \in \{1, \dots, \mu\}$, where $\mu$ is the number
of \emph{meeting groups}, a concept we introduce later on.

\paragraph{Describing the Permutations.}
To describe a solution, we start with the sequence of permutations $\Pi =
[\pi_1, \dots, \pi_\lambda]$.
Each permutation $\pi_r$ is represented by Boolean variables of type $x_{ij}^r$.
These variables describe the relative order of the characters in the
permutation.
The truth assignment of variable $x_{ij}^r$ indicates whether character $i$ is above
character $j$ in permutation $\pi_r$.
To handle ``dead'' characters, we introduce another set of variables $o_i^r$.
Character $i$ is omitted in permutation $\pi_r$ if and only if $o_i^r$
is true.
The clauses described here and under under the following two
headers (constraints for permutations, crossings between
characters, and block crossings) are only active if all involved characters are available (that is, $\neg o_i^r$)
in the permutation considered.
We model this by adding $o_i^r$ as a positive literal to each clause for each
affected character $i$.

To ensure that the variables describe a permutation, we add the following clauses.
We guarantee antisymmetry by $x_{ij}^r \Leftrightarrow
\neg x_{ji}^r$.
We ensure transitivity by
$x_{ij}^r \lor x_{jk}^r \lor x_{ki}^r$ and
$\neg x_{ij}^r \lor \neg x_{jk}^r \lor \neg x_{ki}^r$;
this forces one of the three variables to have a different value than the others.

\paragraph{Crossings Between Characters.}
To simplify the treatment of crossings, we introduce variables that indicate when they occur.
For $r \in \{1, \dots, \lambda - 1\}$, variable~$\chi_{ij}^r$
encodes whether characters~$i$ and~$j$ have a
crossing between permutations~$r$ and $r+1$.
This is the case precisely if they change their relative order between the two
permutations, that is:
$\chi_{ij}^r \Leftrightarrow (x_{ij}^r \neq x_{ij}^{r+1})$ for all
$r \in \{1, \dots, \lambda - 1\}$.
Note that this~-- together with the previously described clauses~--
implies $\chi_{ij}^r \Leftrightarrow \chi_{ji}^r$.
(Recall that constraints involving omitted characters are ``switched off'' using the variables of type~$o_i^r$.)

According to our problem definition we have to ensure that if there is
an addition or removal of characters between
successive permutations, then there can be no block crossing.
So we forbid crossings for all pairs of characters between permutation $\pi_r$
and $\pi_{r+1}$ if a character $i$ is added or removed between these
permutations: $(o_i^r \neq o_i^{r+1}) \Rightarrow \neg \chi_{jk}^r$ for all
$r \in \{1, \dots, \lambda -1\}$.

\paragraph{Block Crossings.}
By the problem definition, there is at most one block
crossing between any two successive permutations $\pi_r$ and $\pi_{r+1}$.
We describe this block crossing by partitioning the character set of permutation
$\pi_r$ into three sets $F_r$, $G_r$, and $H_r$.
For simplicity, we drop the subscript~$r$ in the following.
We express the membership of a character~$i$ in any of these sets using variables $f_i^r$, $g_i^r$ and $h_i^r$, respectively.
Let~$G$ and~$H$ be the two sets of characters that are involved in the potential block crossing between~$\pi_r$ and~$\pi_{r+1}$, and let~$F$ be the set of characters that are not affected by the crossing.
If there is no block crossing between the two permutations, at least one of the two sets~$G$ and~$H$ is empty.

First we add clauses that ensure that every character is in one of the three
sets, that is, exactly one of the variables~$f_i^r$, $g_i^r$, and~$h_i^r$ is
true.
Next, the characters of~$G$ and the characters of~$H$ must each form a contiguous block.
We enforce this by requiring that a character~$j$ is in~$G$ if~$j$ lies
between two characters~$i$ and~$k$ in~$G$:
$x_{ij}^r \land x_{jk}^r \land g_i^r \land g_k^r \Rightarrow g_j^r$.
Similarly, for $H$ we require
$x_{ij}^r \land x_{jk}^r \land h_i^r \land h_k^r \Rightarrow h_j^r$.

We ensure that the blocks~$G$ and~$H$ are adjacent, by requiring that
no character in~$F$ lies between characters in $G$ and in $H$:
$x_{ij}^r \land x_{jk}^r \land g_i^r \land h_k^r \Rightarrow \neg f_j^r$.
Additionally, we prescribe the order of the blocks~$G$ and~$H$ in
the permutation by restricting the characters in~$G$ to be above the characters of~$H$:
$g_i^r \land h_j^r \Rightarrow x_{ij}^r$.

Finally, we ensure that two characters cross each other if and only if they participate in the block crossings, that is, if one
of the characters is in $G$ and the other is in $H$:
$g_i^r \land h_j^r \Leftrightarrow \chi_{ij}^r$ for all
$r \in \{1, \dots, \lambda - 1\}$.

\paragraph{Meeting Groups.}
So far we have introduced various structural constraints to our variables, but we haven't yet established the connection to
our input storyline~$\CS$.
We implement this connection now through the concept of \emph{meeting groups}.
A meeting group is a set of meetings that contain a common point in time.
Instead of the meeting triples (that is, a set of characters, start time, and end time),
we only consider the character sets for the meeting group.
Characters who are alive at that time, but are not part of any meeting, are
added to the meeting group as a singleton meeting.\footnote{This concept of meeting groups is similar to the trees constructed by
Gronemann et al.~\cite{Gronemann2016} to generate MLCM-TC
instances.}
We transform the storyline $\CS$ to a sequence of meeting groups $\CM =
[\CM_1, \dots, \CM_\mu]$ by sorting the events in $\CE$ and putting together
the meetings and live characters for each event in the correct order.
We use $\CM$ only to construct our SAT instance; afterward we transform the
satisfying assignment back into a solution for~$\CS$.

We add variables that connect these meeting groups to the permutations of the
solution.
The variable $q_\ell^r$ indicates whether the meeting group $\CM_\ell$ is assigned to
permutation $\pi_r$.
We require that every meeting group is assigned to
exactly one permutation, that is,
every group is assigned somewhere
$(\bigvee_{r=1}^\lambda q_\ell^r)$ and
no group is assigned twice
$(\neg (q_\ell^r \land q_\ell^p))$.

The meeting groups must be assigned to permutations in the correct order.
If we map $\CM_\ell$ to $\pi_r$, $\CM_{\ell-1}$ has to be assigned to the same
permutation or an earlier one:
$q_\ell^r \Rightarrow \bigvee_{j=1}^r q_{\ell-1}^j$
for $\ell \in \{2, \dots, \mu\}$.
We can assume that the first meeting group is assigned to the first
permutation, as it is not optimal to use block crossings before the
first meetings.
Therefore, we set $q_1^1$ to true.

Next, we handle the birth and death of characters.
Let $\CL_i$ be the meeting groups that contain character $i$.
A permutation $\pi_r$ should contain exactly the characters that are contained
in the assigned meeting groups: those are precisely the alive characters.
We add the clause $q_\ell^r \Rightarrow \neg o_i^r$ if $\CM_\ell \in \CL_i$ and the clause $q_\ell^r \Rightarrow o_i^r$ if $\CM_\ell \notin \CL_i$.
This makes sure that characters involved in meetings must be present and dead characters are omitted.

Note that we allow permutations to not have any meeting groups assigned to them.
This is necessary, for example to allow multiple block crossings
between successive meetings (which may be necessary in an optimal
drawing~\cite{dfflmrsw-bcsv-JGAA17}).
However, such ``loose'' permutations can be exploited to avoid block
crossings by omitting all characters for one permutation and reintroducing them afterward in
an arbitrary order.
To forbid this, for $r = 2, \dots, \lambda$, if no meeting group is assigned to permutation~$\pi_r$,
we do not allow characters to be removed or added in~$\pi_r$:
$\bigwedge_{\ell = 1}^\mu \neg q_\ell^r \Rightarrow (o_i^{r} = o_i^{r-1})$.

Finally, we come to the actual storyline visualization constraint: characters in a meeting must form a contiguous group in the
corresponding permutation.
We add clauses that prohibit characters that are not part of a meeting from
being between characters in the meeting.
That is, if characters $i$ and $k$ are part of a certain meeting in $\CM_\ell$ and $j$ is not, we
have $q_\ell^r \Rightarrow (x_{ij}^r = x_{kj}^r)$.

This concludes our SAT formulation.
If the resulting formula has a satisfying assignment, a solution to our storyline block crossing minimization problem exists, and it is easy to extract the
permutations.
To get the function $A$ that maps the time to the permutations, we have to
remember which meeting group corresponds to which point in time.

Counting the quantifiers in the above construction shows that there are $O(\lambda (\kappa^2 + \mu))$
variables and $O(\lambda \mu (\lambda \kappa^3))$ clauses.
The conjunctive normal form of this SAT formula can clearly be constructed from the storyline in polynomial
time.

\section{Experimental Evaluation}
\label{sec:experiments}

We refer to the approach from Section~\ref{sec:sat} as \aSAT{}.
Additionally, we have implemented two exponential-time exact algorithms that minimize block crossings~\cite{dfflmrsw-bcsv-GD16}.
The first is a branching algorithm that searches for the shortest sequence of block crossings using iterative deepening depth-first search (\aID{}).
This search strategy ensures low memory usage.
The second algorithm is fixed-parameter tractable in the number of characters and works by performing a breadth-first search in an exponentially-large state graph.
Note that these two algorithms do not support concurrent meetings, whereas \aSAT{} does.
We also consider an algorithm by Gronemann et al.~\cite{Gronemann2016} that optimizes pairwise crossings.

\paragraph{Implementation Details.}

All implementations are written in \cpp{}, with the exception of some ``driver'' code in Python for \aSAT{}.
Comparable effort has been put into optimizing each program.
Memory usage was not optimized, but there are no flagrant memory inefficiencies.

\aSAT{} uses the SAT formulation from Section~\ref{sec:sat} and performs exponential search on $\lambda$.
We use \python{} to write CNF SAT instances in DIMACS format, to run \minisat{}~\cite{een2007applying,minisat} on these instances, and to perform the search; the exponential search uses factor 2.
We have used version \mbox{2.2.0} of \minisat{}.\footnote{Slightly modified to measure peak memory usage on Windows.}.
As runtime of \aSAT{}, we report the total time spent by \minisat{}.
This includes all ``real'' work, as well as launching \minisat{} for each formula and the time it spends reading the DIMACS files;
it does not include the runtime of our \python{} code, which has unnecessarily-poor performance and would be unfair in comparison to the other algorithms.

\aID{} and \aDP{} are implemented in \cpp{} following the description in~\cite{dfflmrsw-bcsv-GD16}, including the data structure for block crossings and checking meetings.
For \aID{}, we branch and ``unbranch'' on a single data structure rather than making copies. %
\aDP{} performs a breadth-first search in a large graph. %
We store the nodes explicitly in a flat array addressed by Lehmer codes~\cite{lehmerCodes}:
this requires $\Theta(\numchars!n)$ space, but enables efficient lookup. %
The edges of the graph are enumerated lazily using the ``forward pointers'' from the original paper.

All runtime experiments have been performed on an Intel\textsuperscript\textregistered{} Core\texttrademark{} i5-2400 CPU at \SI{3.10}{\giga\hertz} with \SI{8}{\giga\byte} of RAM and running Windows 7.
This configuration is in some contrast to the experimental setup of Gronemann~et~al.: a $2 \times 10$-core machine with \SI{128}{\giga\byte} of RAM.
Our implementations are single-threaded; their implementation, being based on CPLEX, presumably makes use of the available cores, but this is not reported explicitly.

\paragraph{Real-World Instances.}

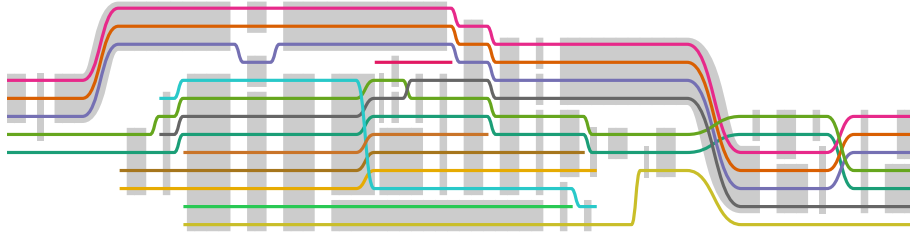
\begin{figure}[tb]
    \centering
    \tikzset{external/export=true}
    \begin{tikzpicture}[scale=0.04]
        \clip (57,0) rectangle (357,88);

        \storylineset{
          xgrid={4,4,4,4,4,4,4,4,4,4,4,4,4,4,4,4,4,4,4,4,16,4,4,4,4,4,4,4,4,4,4,4,4,4,4,4,4,4,4,4,8,4,4,4,4,4,4,4,4,4,4,4,4,4,4,4,4,4,4,4,4,4,4,4,4,4,24,4,4,4,4,4,4,12},
          disty=6,
          meetingaddy=2.2,
        }

        \drawstorylineintervalmeeting{4}{5,5,5,x,x,x,x,x,x,x,x,x,x,x,x,x,x,x,x,x,x,x,x,x,x,x,x,x,x,x,x,x,x,x,x,x,x,x,x,x,x,x,x,x,x,x,x,x,x,x,x,x,x,x,x,x,x,x,x,x,x,x,x,x,x,x,x,x,x,x,x,x,x,x,x,x,x,x,x,x,x,x,x,x,x,x,x,x,x,x,x,x,x,x,x,x,x,x,x,x}
        \drawstorylineintervalmeeting{3}{x,x,x,7,7,7,7,7,7,7,7,7,7,7,7,7,x,x,x,x,x,x,x,x,x,x,x,x,x,x,x,x,x,x,x,x,x,x,x,x,x,x,x,x,x,x,x,x,x,x,x,x,x,x,x,x,x,x,x,x,x,x,x,x,x,x,x,x,x,x,x,x,x,x,x,x,x,x,x,x,x,x,x,x,x,x,x,x,x,x,x,x,x,x,x,x,x,x,x,x}
        \drawstorylineintervalmeeting{2}{x,x,x,x,x,x,x,x,x,x,5,5,5,5,x,x,x,x,x,x,x,x,x,x,x,x,x,x,x,x,x,x,x,x,x,x,x,x,x,x,x,x,x,x,x,x,x,x,x,x,x,x,x,x,x,x,x,x,x,x,x,x,x,x,x,x,x,x,x,x,x,x,x,x,x,x,x,x,x,x,x,x,x,x,x,x,x,x,x,x,x,x,x,x,x,x,x,x,x,x}
        \drawstorylineintervalmeeting{4}{x,x,x,x,x,x,x,x,x,x,x,x,x,x,x,x,6,x,x,x,x,x,x,x,x,x,x,x,x,x,x,x,x,x,x,x,x,x,x,x,x,x,x,x,x,x,x,x,x,x,x,x,x,x,x,x,x,x,x,x,x,x,x,x,x,x,x,x,x,x,x,x,x,x,x,x,x,x,x,x,x,x,x,x,x,x,x,x,x,x,x,x,x,x,x,x,x,x,x,x}
        \drawstorylineintervalmeeting{3}{x,x,x,x,x,x,x,x,x,x,x,x,x,x,x,x,x,7,7,7,11,11,11,11,11,11,11,11,11,11,x,x,x,x,x,x,x,x,x,x,x,x,x,x,x,x,x,x,x,x,x,x,x,x,x,x,x,x,x,x,x,x,x,x,x,x,x,x,x,x,x,x,x,x,x,x,x,x,x,x,x,x,x,x,x,x,x,x,x,x,x,x,x,x,x,x,x,x,x,x}
        \drawstorylineintervalmeeting{4}{x,x,x,x,x,x,x,x,x,x,x,x,x,x,x,x,x,x,x,x,3,3,3,x,x,x,x,x,x,x,x,x,x,x,x,x,x,x,x,x,x,x,x,x,x,x,x,x,x,x,x,x,x,x,x,x,x,x,x,x,x,x,x,x,x,x,x,x,x,x,x,x,x,x,x,x,x,x,x,x,x,x,x,x,x,x,x,x,x,x,x,x,x,x,x,x,x,x,x,x}
        \drawstorylineintervalmeeting{6}{x,x,x,x,x,x,x,x,x,x,x,x,x,x,x,x,x,x,x,x,x,x,x,3,3,x,x,x,x,x,x,x,x,x,x,x,x,x,x,x,x,x,x,x,x,x,x,x,x,x,x,x,x,x,x,x,x,x,x,x,x,x,x,x,x,x,x,x,x,x,x,x,x,x,x,x,x,x,x,x,x,x,x,x,x,x,x,x,x,x,x,x,x,x,x,x,x,x,x,x}
        \drawstorylineintervalmeeting{9}{x,x,x,x,x,x,x,x,x,x,x,x,x,x,x,x,x,x,x,x,x,x,x,x,x,1,1,1,1,1,x,x,x,x,x,x,x,x,x,x,x,x,x,x,x,x,x,x,x,x,x,x,x,x,x,x,x,x,x,x,x,x,x,x,x,x,x,x,x,x,x,x,x,x,x,x,x,x,x,x,x,x,x,x,x,x,x,x,x,x,x,x,x,x,x,x,x,x,x,x}
        \drawstorylineintervalmeeting{8}{x,x,x,x,x,x,x,x,x,x,x,x,x,x,x,x,x,x,x,x,x,x,x,x,x,x,x,x,x,x,1,1,1,x,x,x,x,x,x,x,x,x,x,x,x,x,x,x,x,x,x,x,x,x,x,x,x,x,x,x,x,x,x,x,x,x,x,x,x,x,x,x,x,x,x,x,x,x,x,x,x,x,x,x,x,x,x,x,x,x,x,x,x,x,x,x,x,x,x,x}
        \drawstorylineintervalmeeting{2}{x,x,x,x,x,x,x,x,x,x,x,x,x,x,x,x,x,x,x,x,x,x,x,x,x,x,x,x,x,x,12,12,12,x,x,x,x,x,x,x,x,x,x,x,x,x,x,x,x,x,x,x,x,x,x,x,x,x,x,x,x,x,x,x,x,x,x,x,x,x,x,x,x,x,x,x,x,x,x,x,x,x,x,x,x,x,x,x,x,x,x,x,x,x,x,x,x,x,x,x}
        \drawstorylineintervalmeeting{2}{x,x,x,x,x,x,x,x,x,x,x,x,x,x,x,x,x,x,x,x,x,x,x,x,x,x,x,x,x,x,9,9,9,x,x,x,x,x,x,x,x,x,x,x,x,x,x,x,x,x,x,x,x,x,x,x,x,x,x,x,x,x,x,x,x,x,x,x,x,x,x,x,x,x,x,x,x,x,x,x,x,x,x,x,x,x,x,x,x,x,x,x,x,x,x,x,x,x,x,x}
        \drawstorylineintervalmeeting{9}{x,x,x,x,x,x,x,x,x,x,x,x,x,x,x,x,x,x,x,x,x,x,x,x,x,x,x,x,x,x,x,x,x,1,1,1,1,x,x,x,x,x,x,x,x,x,x,x,x,x,x,x,x,x,x,x,x,x,x,x,x,x,x,x,x,x,x,x,x,x,x,x,x,x,x,x,x,x,x,x,x,x,x,x,x,x,x,x,x,x,x,x,x,x,x,x,x,x,x,x}
        \drawstorylineintervalmeeting{3}{x,x,x,x,x,x,x,x,x,x,x,x,x,x,x,x,x,x,x,x,x,x,x,x,x,x,x,x,x,x,x,x,x,11,11,11,11,11,11,11,11,11,11,11,11,11,11,x,x,x,x,x,x,x,x,x,x,x,x,x,x,x,x,x,x,x,x,x,x,x,x,x,x,x,x,x,x,x,x,x,x,x,x,x,x,x,x,x,x,x,x,x,x,x,x,x,x,x,x,x}
        \drawstorylineintervalmeeting{7}{x,x,x,x,x,x,x,x,x,x,x,x,x,x,x,x,x,x,x,x,x,x,x,x,x,x,x,x,x,x,x,x,x,x,x,x,x,3,3,3,3,x,x,x,x,x,x,x,x,x,x,x,x,x,x,x,x,x,x,x,x,x,x,x,x,x,x,x,x,x,x,x,x,x,x,x,x,x,x,x,x,x,x,x,x,x,x,x,x,x,x,x,x,x,x,x,x,x,x,x}
        \drawstorylineintervalmeeting{2}{x,x,x,x,x,x,x,x,x,x,x,x,x,x,x,x,x,x,x,x,x,x,x,x,x,x,x,x,x,x,x,x,x,x,x,x,x,1,1,1,1,1,1,1,1,1,1,1,1,1,1,1,1,1,1,x,x,x,x,x,x,x,x,x,x,x,x,x,x,x,x,x,x,x,x,x,x,x,x,x,x,x,x,x,x,x,x,x,x,x,x,x,x,x,x,x,x,x,x,x}
        \drawstorylineintervalmeeting[moveinstantaneousright]{3}{x,x,x,x,x,x,x,x,x,x,x,x,x,x,x,x,x,x,x,x,x,x,x,x,x,x,x,x,x,x,x,x,x,x,x,x,x,x,x,x,7,x,x,x,x,x,x,x,x,x,x,x,x,x,x,x,x,x,x,x,x,x,x,x,x,x,x,x,x,x,x,x,x,x,x,x,x,x,x,x,x,x,x,x,x,x,x,x,x,x,x,x,x,x,x,x,x,x,x,x}
        \drawstorylineintervalmeeting{4}{x,x,x,x,x,x,x,x,x,x,x,x,x,x,x,x,x,x,x,x,x,x,x,x,x,x,x,x,x,x,x,x,x,x,x,x,x,x,x,x,3,3,3,3,3,x,x,x,x,x,x,x,x,x,x,x,x,x,x,x,x,x,x,x,x,x,x,x,x,x,x,x,x,x,x,x,x,x,x,x,x,x,x,x,x,x,x,x,x,x,x,x,x,x,x,x,x,x,x,x}
        \drawstorylineintervalmeeting{2}{x,x,x,x,x,x,x,x,x,x,x,x,x,x,x,x,x,x,x,x,x,x,x,x,x,x,x,x,x,x,x,x,x,x,x,x,x,x,x,x,x,7,7,x,x,x,x,x,x,x,x,x,x,x,x,x,x,x,x,x,x,x,x,x,x,x,x,x,x,x,x,x,x,x,x,x,x,x,x,x,x,x,x,x,x,x,x,x,x,x,x,x,x,x,x,x,x,x,x,x}
        \drawstorylineintervalmeeting{2}{x,x,x,x,x,x,x,x,x,x,x,x,x,x,x,x,x,x,x,x,x,x,x,x,x,x,x,x,x,x,x,x,x,x,x,x,x,x,x,x,x,9,9,x,x,x,x,x,x,x,x,x,x,x,x,x,x,x,x,x,x,x,x,x,x,x,x,x,x,x,x,x,x,x,x,x,x,x,x,x,x,x,x,x,x,x,x,x,x,x,x,x,x,x,x,x,x,x,x,x}
        \drawstorylineintervalmeeting{3}{x,x,x,x,x,x,x,x,x,x,x,x,x,x,x,x,x,x,x,x,x,x,x,x,x,x,x,x,x,x,x,x,x,x,x,x,x,x,x,x,x,x,x,7,7,x,x,x,x,x,x,x,x,x,x,x,x,x,x,x,x,x,x,x,x,x,x,x,x,x,x,x,x,x,x,x,x,x,x,x,x,x,x,x,x,x,x,x,x,x,x,x,x,x,x,x,x,x,x,x}
        \drawstorylineintervalmeeting{7}{x,x,x,x,x,x,x,x,x,x,x,x,x,x,x,x,x,x,x,x,x,x,x,x,x,x,x,x,x,x,x,x,x,x,x,x,x,x,x,x,x,x,x,x,x,3,3,x,x,x,x,x,x,x,x,x,x,x,x,x,x,x,x,x,x,x,x,x,x,x,x,x,x,x,x,x,x,x,x,x,x,x,x,x,x,x,x,x,x,x,x,x,x,x,x,x,x,x,x,x}
        \drawstorylineintervalmeeting{10}{x,x,x,x,x,x,x,x,x,x,x,x,x,x,x,x,x,x,x,x,x,x,x,x,x,x,x,x,x,x,x,x,x,x,x,x,x,x,x,x,x,x,x,x,x,x,x,3,3,3,x,x,x,x,x,x,x,x,x,x,x,x,x,x,x,x,x,x,x,x,x,x,x,x,x,x,x,x,x,x,x,x,x,x,x,x,x,x,x,x,x,x,x,x,x,x,x,x,x,x}
        \drawstorylineintervalmeeting{9}{x,x,x,x,x,x,x,x,x,x,x,x,x,x,x,x,x,x,x,x,x,x,x,x,x,x,x,x,x,x,x,x,x,x,x,x,x,x,x,x,x,x,x,x,x,x,x,x,x,x,3,3,3,x,x,x,x,x,x,x,x,x,x,x,x,x,x,x,x,x,x,x,x,x,x,x,x,x,x,x,x,x,x,x,x,x,x,x,x,x,x,x,x,x,x,x,x,x,x,x}
        \drawstorylineintervalmeeting{5}{x,x,x,x,x,x,x,x,x,x,x,x,x,x,x,x,x,x,x,x,x,x,x,x,x,x,x,x,x,x,x,x,x,x,x,x,x,x,x,x,x,x,x,x,x,x,x,x,x,x,x,x,x,3,3,x,x,x,x,x,x,x,x,x,x,x,x,x,x,x,x,x,x,x,x,x,x,x,x,x,x,x,x,x,x,x,x,x,x,x,x,x,x,x,x,x,x,x,x,x}
        \drawstorylineintervalmeeting{2}{x,x,x,x,x,x,x,x,x,x,x,x,x,x,x,x,x,x,x,x,x,x,x,x,x,x,x,x,x,x,x,x,x,x,x,x,x,x,x,x,x,x,x,x,x,x,x,x,x,x,x,x,x,10,10,x,x,x,x,x,x,x,x,x,x,x,x,x,x,x,x,x,x,x,x,x,x,x,x,x,x,x,x,x,x,x,x,x,x,x,x,x,x,x,x,x,x,x,x,x}
        \drawstorylineintervalmeeting{2}{x,x,x,x,x,x,x,x,x,x,x,x,x,x,x,x,x,x,x,x,x,x,x,x,x,x,x,x,x,x,x,x,x,x,x,x,x,x,x,x,x,x,x,x,x,x,x,x,x,x,x,x,x,8,8,x,x,x,x,x,x,x,x,x,x,x,x,x,x,x,x,x,x,x,x,x,x,x,x,x,x,x,x,x,x,x,x,x,x,x,x,x,x,x,x,x,x,x,x,x}
        \drawstorylineintervalmeeting{4}{x,x,x,x,x,x,x,x,x,x,x,x,x,x,x,x,x,x,x,x,x,x,x,x,x,x,x,x,x,x,x,x,x,x,x,x,x,x,x,x,x,x,x,x,x,x,x,x,x,x,x,x,x,x,x,4,4,4,x,x,x,x,x,x,x,x,x,x,x,x,x,x,x,x,x,x,x,x,x,x,x,x,x,x,x,x,x,x,x,x,x,x,x,x,x,x,x,x,x,x}
        \drawstorylineintervalmeeting{4}{x,x,x,x,x,x,x,x,x,x,x,x,x,x,x,x,x,x,x,x,x,x,x,x,x,x,x,x,x,x,x,x,x,x,x,x,x,x,x,x,x,x,x,x,x,x,x,x,x,x,x,x,x,x,x,8,8,8,8,8,8,8,8,8,8,8,2,2,x,x,x,x,x,x,x,x,x,x,x,x,x,x,x,x,x,x,x,x,x,x,x,x,x,x,x,x,x,x,x,x}
        \drawstorylineintervalmeeting{3}{x,x,x,x,x,x,x,x,x,x,x,x,x,x,x,x,x,x,x,x,x,x,x,x,x,x,x,x,x,x,x,x,x,x,x,x,x,x,x,x,x,x,x,x,x,x,x,x,x,x,x,x,x,x,x,1,1,x,x,x,x,x,x,x,x,x,x,x,x,x,x,x,x,x,x,x,x,x,x,x,x,x,x,x,x,x,x,x,x,x,x,x,x,x,x,x,x,x,x,x}
        \drawstorylineintervalmeeting{2}{x,x,x,x,x,x,x,x,x,x,x,x,x,x,x,x,x,x,x,x,x,x,x,x,x,x,x,x,x,x,x,x,x,x,x,x,x,x,x,x,x,x,x,x,x,x,x,x,x,x,x,x,x,x,x,x,x,1,1,x,x,x,x,x,x,x,x,x,x,x,x,x,x,x,x,x,x,x,x,x,x,x,x,x,x,x,x,x,x,x,x,x,x,x,x,x,x,x,x,x}
        \drawstorylineintervalmeeting{3}{x,x,x,x,x,x,x,x,x,x,x,x,x,x,x,x,x,x,x,x,x,x,x,x,x,x,x,x,x,x,x,x,x,x,x,x,x,x,x,x,x,x,x,x,x,x,x,x,x,x,x,x,x,x,x,x,x,x,4,x,x,x,x,x,x,x,x,x,x,x,x,x,x,x,x,x,x,x,x,x,x,x,x,x,x,x,x,x,x,x,x,x,x,x,x,x,x,x,x,x}
        \drawstorylineintervalmeeting{2}{x,x,x,x,x,x,x,x,x,x,x,x,x,x,x,x,x,x,x,x,x,x,x,x,x,x,x,x,x,x,x,x,x,x,x,x,x,x,x,x,x,x,x,x,x,x,x,x,x,x,x,x,x,x,x,x,x,x,x,5,5,5,x,x,x,x,x,x,x,x,x,x,x,x,x,x,x,x,x,x,x,x,x,x,x,x,x,x,x,x,x,x,x,x,x,x,x,x,x,x}
        \drawstorylineintervalmeeting[moveinstantaneousright]{2}{x,x,x,x,x,x,x,x,x,x,x,x,x,x,x,x,x,x,x,x,x,x,x,x,x,x,x,x,x,x,x,x,x,x,x,x,x,x,x,x,x,x,x,x,x,x,x,x,x,x,x,x,x,x,x,x,x,x,x,x,x,x,4,x,x,x,x,x,x,x,x,x,x,x,x,x,x,x,x,x,x,x,x,x,x,x,x,x,x,x,x,x,x,x,x,x,x,x,x,x}
        \drawstorylineintervalmeeting{3}{x,x,x,x,x,x,x,x,x,x,x,x,x,x,x,x,x,x,x,x,x,x,x,x,x,x,x,x,x,x,x,x,x,x,x,x,x,x,x,x,x,x,x,x,x,x,x,x,x,x,x,x,x,x,x,x,x,x,x,x,x,x,x,4,4,4,x,x,x,x,x,x,x,x,x,x,x,x,x,x,x,x,x,x,x,x,x,x,x,x,x,x,x,x,x,x,x,x,x,x}
        \drawstorylineintervalmeeting{2}{x,x,x,x,x,x,x,x,x,x,x,x,x,x,x,x,x,x,x,x,x,x,x,x,x,x,x,x,x,x,x,x,x,x,x,x,x,x,x,x,x,x,x,x,x,x,x,x,x,x,x,x,x,x,x,x,x,x,x,x,x,x,x,x,x,x,6,6,x,x,x,x,x,x,x,x,x,x,x,x,x,x,x,x,x,x,x,x,x,x,x,x,x,x,x,x,x,x,x,x}
        \drawstorylineintervalmeeting{3}{x,x,x,x,x,x,x,x,x,x,x,x,x,x,x,x,x,x,x,x,x,x,x,x,x,x,x,x,x,x,x,x,x,x,x,x,x,x,x,x,x,x,x,x,x,x,x,x,x,x,x,x,x,x,x,x,x,x,x,x,x,x,x,x,x,x,x,x,5,5,5,x,x,x,x,x,x,x,x,x,x,x,x,x,x,x,x,x,x,x,x,x,x,x,x,x,x,x,x,x}
        \drawstorylineintervalmeeting{3}{x,x,x,x,x,x,x,x,x,x,x,x,x,x,x,x,x,x,x,x,x,x,x,x,x,x,x,x,x,x,x,x,x,x,x,x,x,x,x,x,x,x,x,x,x,x,x,x,x,x,x,x,x,x,x,x,x,x,x,x,x,x,x,x,x,x,x,x,2,2,2,2,x,x,x,x,x,x,x,x,x,x,x,x,x,x,x,x,x,x,x,x,x,x,x,x,x,x,x,x}
        \drawstorylineintervalmeeting{2}{x,x,x,x,x,x,x,x,x,x,x,x,x,x,x,x,x,x,x,x,x,x,x,x,x,x,x,x,x,x,x,x,x,x,x,x,x,x,x,x,x,x,x,x,x,x,x,x,x,x,x,x,x,x,x,x,x,x,x,x,x,x,x,x,x,x,x,x,x,x,x,6,6,x,x,x,x,x,x,x,x,x,x,x,x,x,x,x,x,x,x,x,x,x,x,x,x,x,x,x}
        \drawstorylineintervalmeeting{4}{x,x,x,x,x,x,x,x,x,x,x,x,x,x,x,x,x,x,x,x,x,x,x,x,x,x,x,x,x,x,x,x,x,x,x,x,x,x,x,x,x,x,x,x,x,x,x,x,x,x,x,x,x,x,x,x,x,x,x,x,x,x,x,x,x,x,x,x,x,x,x,x,2,x,x,x,x,x,x,x,x,x,x,x,x,x,x,x,x,x,x,x,x,x,x,x,x,x,x,x}
        \drawstorylineintervalmeeting{6}{x,x,x,x,x,x,x,x,x,x,x,x,x,x,x,x,x,x,x,x,x,x,x,x,x,x,x,x,x,x,x,x,x,x,x,x,x,x,x,x,x,x,x,x,x,x,x,x,x,x,x,x,x,x,x,x,x,x,x,x,x,x,x,x,x,x,x,x,x,x,x,x,x,2,2,x,x,x,x,x,x,x,x,x,x,x,x,x,x,x,x,x,x,x,x,x,x,x,x,x}
        \drawstorylineintervalmeeting{3}{x,x,x,x,x,x,x,x,x,x,x,x,x,x,x,x,x,x,x,x,x,x,x,x,x,x,x,x,x,x,x,x,x,x,x,x,x,x,x,x,x,x,x,x,x,x,x,x,x,x,x,x,x,x,x,x,x,x,x,x,x,x,x,x,x,x,x,x,x,x,x,x,x,x,x,2,2,2,2,x,x,x,x,x,x,x,x,x,x,x,x,x,x,x,x,x,x,x,x,x}
        \drawstorylineintervalmeeting{3}{x,x,x,x,x,x,x,x,x,x,x,x,x,x,x,x,x,x,x,x,x,x,x,x,x,x,x,x,x,x,x,x,x,x,x,x,x,x,x,x,x,x,x,x,x,x,x,x,x,x,x,x,x,x,x,x,x,x,x,x,x,x,x,x,x,x,x,x,x,x,x,x,x,x,x,x,5,5,5,5,5,x,x,x,x,x,x,x,x,x,x,x,x,x,x,x,x,x,x,x}
        \drawstorylineintervalmeeting{2}{x,x,x,x,x,x,x,x,x,x,x,x,x,x,x,x,x,x,x,x,x,x,x,x,x,x,x,x,x,x,x,x,x,x,x,x,x,x,x,x,x,x,x,x,x,x,x,x,x,x,x,x,x,x,x,x,x,x,x,x,x,x,x,x,x,x,x,x,x,x,x,x,x,x,x,x,x,x,x,3,3,x,x,x,x,x,x,x,x,x,x,x,x,x,x,x,x,x,x,x}
        \drawstorylineintervalmeeting{2}{x,x,x,x,x,x,x,x,x,x,x,x,x,x,x,x,x,x,x,x,x,x,x,x,x,x,x,x,x,x,x,x,x,x,x,x,x,x,x,x,x,x,x,x,x,x,x,x,x,x,x,x,x,x,x,x,x,x,x,x,x,x,x,x,x,x,x,x,x,x,x,x,x,x,x,x,x,x,x,1,1,x,x,x,x,x,x,x,x,x,x,x,x,x,x,x,x,x,x,x}
        \drawstorylineintervalmeeting{3}{x,x,x,x,x,x,x,x,x,x,x,x,x,x,x,x,x,x,x,x,x,x,x,x,x,x,x,x,x,x,x,x,x,x,x,x,x,x,x,x,x,x,x,x,x,x,x,x,x,x,x,x,x,x,x,x,x,x,x,x,x,x,x,x,x,x,x,x,x,x,x,x,x,x,x,x,x,x,x,x,x,1,1,1,1,1,1,1,1,1,1,1,1,1,1,1,x,x,x,x}
        \drawstorylineintervalmeeting{2}{x,x,x,x,x,x,x,x,x,x,x,x,x,x,x,x,x,x,x,x,x,x,x,x,x,x,x,x,x,x,x,x,x,x,x,x,x,x,x,x,x,x,x,x,x,x,x,x,x,x,x,x,x,x,x,x,x,x,x,x,x,x,x,x,x,x,x,x,x,x,x,x,x,x,x,x,x,x,x,x,x,6,6,6,6,6,x,x,x,x,x,x,x,x,x,x,x,x,x,x}
        \drawstorylineintervalmeeting{2}{x,x,x,x,x,x,x,x,x,x,x,x,x,x,x,x,x,x,x,x,x,x,x,x,x,x,x,x,x,x,x,x,x,x,x,x,x,x,x,x,x,x,x,x,x,x,x,x,x,x,x,x,x,x,x,x,x,x,x,x,x,x,x,x,x,x,x,x,x,x,x,x,x,x,x,x,x,x,x,x,x,4,4,4,4,4,x,x,x,x,x,x,x,x,x,x,x,x,x,x}
        \drawstorylineintervalmeeting{4}{x,x,x,x,x,x,x,x,x,x,x,x,x,x,x,x,x,x,x,x,x,x,x,x,x,x,x,x,x,x,x,x,x,x,x,x,x,x,x,x,x,x,x,x,x,x,x,x,x,x,x,x,x,x,x,x,x,x,x,x,x,x,x,x,x,x,x,x,x,x,x,x,x,x,x,x,x,x,x,x,x,x,x,x,x,x,4,4,x,x,x,x,x,x,x,x,x,x,x,x}
        \drawstorylineintervalmeeting{2}{x,x,x,x,x,x,x,x,x,x,x,x,x,x,x,x,x,x,x,x,x,x,x,x,x,x,x,x,x,x,x,x,x,x,x,x,x,x,x,x,x,x,x,x,x,x,x,x,x,x,x,x,x,x,x,x,x,x,x,x,x,x,x,x,x,x,x,x,x,x,x,x,x,x,x,x,x,x,x,x,x,x,x,x,x,x,x,x,6,6,6,6,6,6,x,x,x,x,x,x}
        \drawstorylineintervalmeeting{2}{x,x,x,x,x,x,x,x,x,x,x,x,x,x,x,x,x,x,x,x,x,x,x,x,x,x,x,x,x,x,x,x,x,x,x,x,x,x,x,x,x,x,x,x,x,x,x,x,x,x,x,x,x,x,x,x,x,x,x,x,x,x,x,x,x,x,x,x,x,x,x,x,x,x,x,x,x,x,x,x,x,x,x,x,x,x,x,x,4,4,x,x,x,x,x,x,x,x,x,x}
        \drawstorylineintervalmeeting{3}{x,x,x,x,x,x,x,x,x,x,x,x,x,x,x,x,x,x,x,x,x,x,x,x,x,x,x,x,x,x,x,x,x,x,x,x,x,x,x,x,x,x,x,x,x,x,x,x,x,x,x,x,x,x,x,x,x,x,x,x,x,x,x,x,x,x,x,x,x,x,x,x,x,x,x,x,x,x,x,x,x,x,x,x,x,x,x,x,x,x,x,x,x,x,5,5,x,x,x,x}
        \drawstorylineintervalmeeting{4}{x,x,x,x,x,x,x,x,x,x,x,x,x,x,x,x,x,x,x,x,x,x,x,x,x,x,x,x,x,x,x,x,x,x,x,x,x,x,x,x,x,x,x,x,x,x,x,x,x,x,x,x,x,x,x,x,x,x,x,x,x,x,x,x,x,x,x,x,x,x,x,x,x,x,x,x,x,x,x,x,x,x,x,x,x,x,x,x,x,x,x,x,x,x,x,x,1,x,x,x}
        \drawstorylineintervalmeeting{2}{x,x,x,x,x,x,x,x,x,x,x,x,x,x,x,x,x,x,x,x,x,x,x,x,x,x,x,x,x,x,x,x,x,x,x,x,x,x,x,x,x,x,x,x,x,x,x,x,x,x,x,x,x,x,x,x,x,x,x,x,x,x,x,x,x,x,x,x,x,x,x,x,x,x,x,x,x,x,x,x,x,x,x,x,x,x,x,x,x,x,x,x,x,x,x,x,6,6,6,6}
        \drawstorylineintervalmeeting{5}{x,x,x,x,x,x,x,x,x,x,x,x,x,x,x,x,x,x,x,x,x,x,x,x,x,x,x,x,x,x,x,x,x,x,x,x,x,x,x,x,x,x,x,x,x,x,x,x,x,x,x,x,x,x,x,x,x,x,x,x,x,x,x,x,x,x,x,x,x,x,x,x,x,x,x,x,x,x,x,x,x,x,x,x,x,x,x,x,x,x,x,x,x,x,x,x,x,0,x,x}
        \drawstorylineintervalmeeting{4}{x,x,x,x,x,x,x,x,x,x,x,x,x,x,x,x,x,x,x,x,x,x,x,x,x,x,x,x,x,x,x,x,x,x,x,x,x,x,x,x,x,x,x,x,x,x,x,x,x,x,x,x,x,x,x,x,x,x,x,x,x,x,x,x,x,x,x,x,x,x,x,x,x,x,x,x,x,x,x,x,x,x,x,x,x,x,x,x,x,x,x,x,x,x,x,x,x,x,1,1}

        \drawstoryline[drawingstyle=char1]{5,5,5,5,5,5,5,5,5,5,5,5,5,5,5,5,5,5,5,5,5,5,5,5,5,6,6,6,6,6,6,6,6,6,6,6,6,6,6,6,7,7,7,7,7,7,7,7,7,7,6,6,6,6,6,6,6,6,5,5,5,5,5,5,5,5,6,6,6,6,6,6,6,3,3,3,3,3,3,3,3,3,3,3,3,3,3,3,3,3,3,3,3,3,3,3,3,3,3,3}
        \drawstoryline[drawingstyle=char2]{7,7,7,8,8,8,8,8,8,8,8,8,8,8,8,8,8,8,8,8,12,12,12,12,12,12,12,12,12,12,12,12,12,12,12,12,12,12,12,12,12,12,12,12,12,12,12,11,11,11,10,10,10,10,10,10,10,10,10,10,10,10,10,10,10,10,4,4,4,4,4,4,4,6,6,6,6,6,6,6,6,6,6,6,6,6,6,6,6,6,6,6,6,6,6,6,6,6,6,6}
        \drawstoryline[drawingstyle=char3]{6,6,6,7,7,7,7,7,7,7,7,7,7,7,7,7,7,7,7,7,11,11,11,11,11,11,11,11,11,11,10,10,10,11,11,11,11,11,11,11,11,11,11,11,11,11,11,10,10,10,9,9,9,9,9,9,9,9,9,9,9,9,9,9,9,9,3,3,3,3,3,3,3,5,5,5,5,5,5,5,5,5,5,5,5,5,5,5,5,5,x,x,x,x,x,x,x,x,x,x}
        \drawstoryline[drawingstyle=char4]{8,8,8,9,9,9,9,9,9,9,9,9,9,9,9,9,9,9,9,9,13,13,13,13,13,13,13,13,13,13,13,13,13,13,13,13,13,13,13,13,13,13,13,13,13,13,13,12,12,12,11,11,11,11,11,11,11,11,11,11,11,11,11,11,11,11,5,5,5,5,5,5,5,7,7,7,7,7,7,7,7,7,7,7,7,7,7,7,7,7,7,7,7,7,7,7,7,7,7,7}
        \drawstoryline[drawingstyle=char5]{x,x,x,x,x,x,6,6,6,6,6,6,6,6,6,6,6,6,6,6,6,6,6,7,7,8,8,8,8,8,8,8,8,8,8,8,8,8,8,8,9,9,9,8,8,8,8,8,8,8,7,7,7,7,7,7,7,7,6,6,6,6,6,6,6,6,7,7,7,7,7,7,7,4,4,4,4,4,4,4,4,4,4,4,4,4,4,4,4,4,4,4,4,4,5,5,4,4,4,4}
        \drawstoryline[drawingstyle=char6]{x,x,x,x,x,x,x,x,x,x,x,x,x,x,x,x,x,x,x,x,3,3,3,3,3,3,3,3,3,3,3,3,3,3,3,3,3,3,3,3,4,4,4,4,4,4,4,4,4,4,4,4,4,4,4,4,4,4,4,x,x,x,x,x,x,x,x,x,x,x,x,x,x,x,x,x,x,x,x,x,x,x,x,x,x,x,x,x,x,x,x,x,x,x,x,x,x,x,x,x}
        \drawstoryline[drawingstyle=char7]{x,x,x,x,x,x,x,x,x,x,x,x,x,x,x,x,x,x,x,x,4,4,4,4,4,4,4,4,4,4,4,4,4,4,4,4,4,4,4,4,5,5,5,5,5,5,5,5,5,5,5,5,5,5,5,5,5,5,x,x,x,x,x,x,x,x,x,x,x,x,x,x,x,x,x,x,x,x,x,x,x,x,x,x,x,x,x,x,x,x,x,x,x,x,x,x,x,x,x,x}
        \drawstoryline[drawingstyle=char8]{x,x,x,x,x,x,x,x,x,x,x,x,x,x,x,x,x,x,x,x,x,x,x,6,6,7,7,7,7,7,7,7,7,7,7,7,7,7,7,7,8,8,8,9,9,9,9,9,9,9,8,8,8,8,8,8,8,8,8,8,8,8,8,8,8,8,2,2,2,2,2,2,2,2,2,2,2,2,2,2,2,2,2,2,2,2,2,2,2,2,2,2,2,2,2,2,2,2,2,2}
        \drawstoryline[drawingstyle=char9]{x,x,x,x,x,x,x,x,x,x,x,x,x,x,x,x,x,x,x,x,x,x,x,8,8,9,9,9,9,9,9,9,9,9,9,9,9,9,9,9,3,3,3,3,3,3,3,3,3,3,3,3,3,3,3,3,3,2,2,x,x,x,x,x,x,x,x,x,x,x,x,x,x,x,x,x,x,x,x,x,x,x,x,x,x,x,x,x,x,x,x,x,x,x,x,x,x,x,x,x}
        \drawstoryline[drawingstyle=char10]{x,x,x,x,x,x,x,x,x,x,x,x,x,x,x,x,x,x,x,x,x,x,x,x,x,2,2,2,2,2,2,2,2,2,2,2,2,2,2,2,2,2,2,2,2,2,2,2,2,2,2,2,2,2,2,2,2,x,x,x,x,x,x,x,x,x,x,x,x,x,x,x,x,x,x,x,x,x,x,x,x,x,x,x,x,x,x,x,x,x,x,x,x,x,x,x,x,x,x,x}
        \drawstoryline[drawingstyle=char11]{x,x,x,x,x,x,x,x,x,x,x,x,x,x,x,x,x,x,x,x,x,x,x,x,x,1,1,1,1,1,1,1,1,1,1,1,1,1,1,1,1,1,1,1,1,1,1,1,1,1,1,1,1,1,1,1,1,1,1,1,1,1,4,4,4,4,1,1,1,1,1,1,1,1,1,1,1,1,1,1,1,1,1,1,1,1,1,1,1,1,1,1,1,1,1,1,1,1,1,1}
        \drawstoryline[drawingstyle=char12]{x,x,x,x,x,x,x,x,x,x,x,x,x,x,x,x,x,x,x,x,x,x,x,x,x,5,5,5,5,5,5,5,5,5,5,5,5,5,5,5,6,6,6,6,6,6,6,6,6,6,x,x,x,x,x,x,x,x,x,x,x,x,x,x,x,x,x,x,x,x,x,x,x,x,x,x,x,x,x,x,x,x,x,x,x,x,x,x,x,x,x,x,x,x,x,x,x,x,x,x}
        \drawstoryline[drawingstyle=char13]{x,x,x,x,x,x,x,x,x,x,x,x,x,x,x,x,x,x,x,x,x,x,x,x,x,x,x,x,x,x,x,x,x,x,x,x,x,x,x,x,10,10,10,10,10,10,10,x,x,x,x,x,x,x,x,x,x,x,x,x,x,x,x,x,x,x,x,x,x,x,x,x,x,x,x,x,x,x,x,x,x,x,x,x,x,x,x,x,x,x,x,x,x,x,x,x,x,x,x,x}
        \drawstoryline[drawingstyle=char14]{x,x,x,x,x,x,x,x,x,x,x,x,x,x,x,x,x,x,x,x,x,x,x,x,x,x,x,x,x,x,x,x,x,x,x,x,x,x,x,x,x,x,x,x,x,x,x,x,x,x,x,x,x,x,x,x,x,x,x,x,x,x,x,x,x,x,x,x,x,x,x,x,x,x,x,x,x,x,x,x,x,x,x,x,x,x,0,0,0,0,0,0,0,0,0,0,0,0,x,x}
    \end{tikzpicture}
    \tikzset{external/export=false}
    \caption{A snippet of a block-crossing optimal drawing of The Matrix based on the sequence of permutations found by \aSAT{}, and the start and end times of the meetings (visualized by the gray blocks).
    The drawing reflects the linear order of the events but not their absolute points in time.
    \label{fig:thematrix}}
\end{figure}

We use the same real-world instances as Gronemann et al.~\cite{Gronemann2016}.
These include three movies
and chapters from several books.
See Figure~\ref{fig:thematrix} for a block-crossing optimal drawing of The Matrix computed using \aSAT{}.
More drawings computed using \aSAT{} are found in Appendix~\ref{app:drawings}.

Table~\ref{tab:movies} compares our block crossing optimization to solutions optimized for pairwise crossings.
It shows that the optimal number of block crossings is much lower than the optimal number of pairwise crossings.
This decrease is not just counting things differently: Gronemann et~al.'s drawing of The Matrix, for example, has the optimal number of 12 crossings and happens to have 8 block crossings, whereas we give an optimal drawing with 4 block crossings.
Our drawing happens to have 33 pairwise crossings: this presents an interesting trade-off.

\begin{table}[t]
\begin{center}
\begin{tabular}{lcccccccc}
\toprule
 & \multicolumn{5}{c}{Block crossings using \aSAT{}} & \multicolumn{3}{c}{Gronemann~et~al.~\cite{Gronemann2016}} \\
\cmidrule(r){2-6}
\cmidrule(r){7-9}
Instance & {\enskip $\lambda_\mathrm{OPT}$\enskip} & {\enskip Memory\enskip} & {\enskip cr\enskip} & {\enskip bc\textsubscript{OPT}\enskip} & {\enskip Time [\si{\second}]\enskip} & {\enskip cr\textsubscript{OPT}\enskip} & {\enskip bc\enskip} & {\enskip Time [\si{\second}]\enskip} \\
\midrule
Star Wars        & 20 & \SI{79}{\mega\byte} & 54 & 10 & 3.77 & 39 & 18 & 0.99 \\
The Matrix \quad & 18 & \SI{67}{\mega\byte} & 21 & 4  & 2.86 & 12 & 8  & 0.77 \\
Inception        & 23 & \SI{39}{\mega\byte} & 51 & 12 & 1.54 & 35 & 20 & 2.02 \\
\bottomrule
\end{tabular}
\end{center}
\caption{Comparison of pairwise crossings (cr) and block crossings (bc) on movie instances; subscript OPT indicates the value that the algorithm optimized.
The runtime of both approaches is similar, even on rather different
machines (see Section~\ref{sec:experiments} -- ``Implementation Details''.).
\label{tab:movies}}
\end{table}

The book instances
unfortunately present a strong challenge for our algorithms.
Even though there are no concurrent meetings, the number of characters immediately disqualifies \aDP{} and the optimum is too large for \aID{}.
This leaves \aSAT{}, but these instances
(as modeled in Gronemann et al.)
contain an extreme number of `births' and `deaths.'
While this is convenient for their algorithm (or at least: not detrimental), our SAT formulation requires a large number of permutations to handle this.
One might hope that~-- even though large~-- these formulas are still
relatively easy for \minisat{}: alas, they are not.

\begin{figure}[t]
\centering
\includegraphics{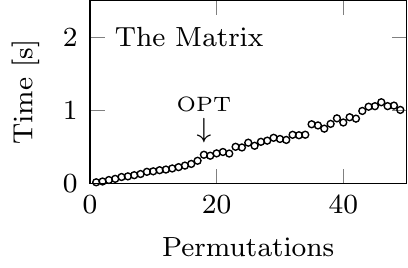}
\hfill
\includegraphics{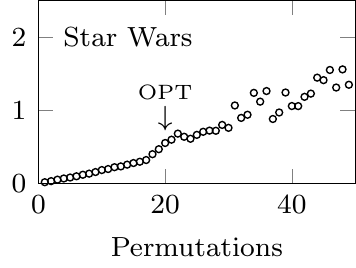}
\hfill
\includegraphics{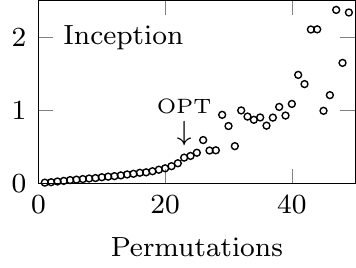}
\caption{Runtime of \minisat{} for different numbers of permutations on the movie instances.
Recall that the number of permutations does not equal the number of block crossings.
\label{fig:sat_try}}
\end{figure}

Finally, we look at the exponential search that \aSAT{} uses to find the optimal number of permutations.
If testing a number of permutations takes exponential time (we are solving a SAT instance, after all), a single overestimate would be disastrous.
However, on the real-world instances we observe fairly modest time for overestimated $\lambda$ (see Figure~\ref{fig:sat_try}).
This means exponential search can have a significant advantage over linear search.
On the movie instances, using exponential search is indeed faster than linear search, but just by about a third.

\paragraph{Random Instances.}

We test using random instances of two kinds.
The first are \emph{uniform} instances and
	these are the same as in previous work~\cite{dfflmrsw-bcsv-GD16}.
	First, pick $\numchars$, $n$, and a probability $p$.
	(We report here on $p=0.5$.)
	Then generate a meeting by picking, independently at random with probability $p$, whether each character is in the meeting.
	Reject meetings with fewer than two characters, and repeat until there are $n$ meetings.
	We let all characters be alive at all times, so we can run all three algorithms.

\begin{figure}[t]
\centering
\includegraphics{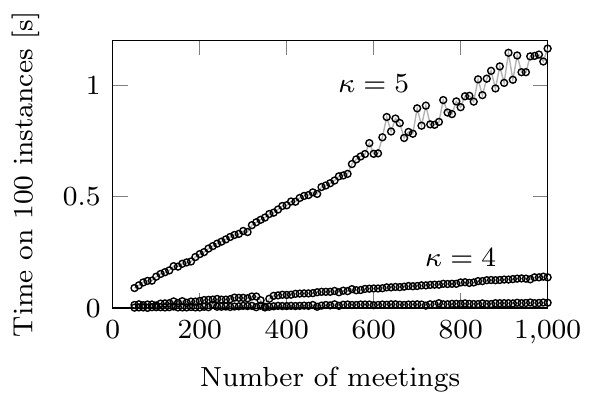}
\hfill
\includegraphics{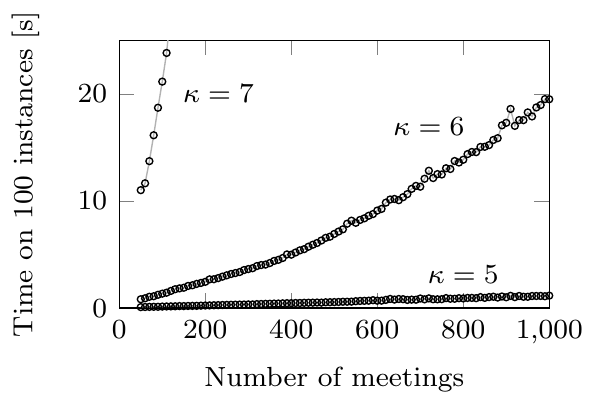}
\caption{Total runtime of \aDP{} on 100 random instances from the uniform model with $p=0.5$, increasing number of meetings, and $\numchars\in\{3,4,5,6,7\}$.
\label{fig:dp_runtime_k}}
\end{figure}

Figure~\ref{fig:dp_runtime_k} shows the runtime of \aDP{} on these instances as a function of $n$, for various numbers of characters.
It confirms the fixed-parameter tractable runtime in practice.
Note that the plot reports the runtime for solving 100 instances.
The other algorithms have trouble handling 1000 meetings in any reasonable setting; with $\numchars=5$, \aDP{} solves 100 such instances in little more than a second.
However, the explosive dependence on $\numchars$ is also clear.

\begin{figure}[t]
\centering
\includegraphics{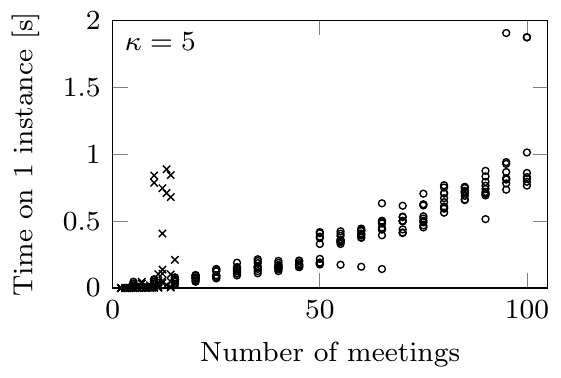}
\hfill
\includegraphics{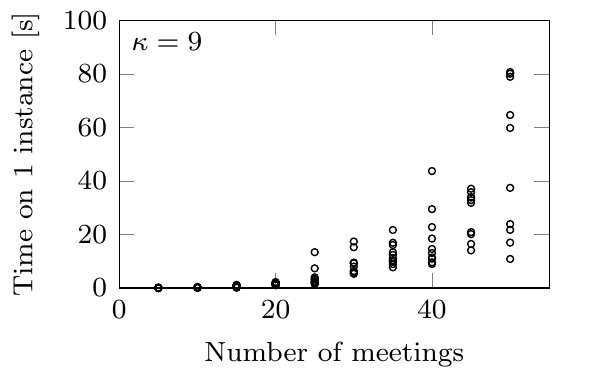}
\caption{Circular marks: runtime of \aSAT{} on instances from the uniform model with $p=0.5$, and $\numchars=5$ (left) and $\numchars=9$ (right).
Crosses, left plot: runtime of \aID{}.
It is highly variable and practically dominated by \aSAT{}.
\label{fig:sat_runtime_uni}}
\end{figure}

Figure~\ref{fig:sat_runtime_uni} similarly shows the runtime of \aSAT{}.
Since there is more variance, we show 10 data points for every number of meetings, rather than the sum.
For $\numchars=5$, \aSAT{} can easily handle 100 meetings: except for an outlier, we are not yet hit by a runtime explosion.
Note, however, that it is significantly slower than \aDP{}:
approximately three orders of magnitude at $n=100$.
As for \aID{}: it is so slow on these instances that its runtime escapes the plot almost immediately.

For $\numchars=9$, a different picture develops.
Firstly, \aSAT{} experiences difficulty as the number of meetings increases.
With instances approaching 50 meetings, the runtime starts to explode.
For these instances, the runtime of \aDP{} is similar since it too has become slow at $\numchars=9$.
The difference is that, if we are willing to wait longer, \aSAT{} can
be run on instances with more than 9 characters, whereas \aDP{} is
quite fundamentally limited by its memory usage (see
``Memory Usage'').

Yet another picture emerges when we look at instances that have a solution with few block crossings.
	To consider such instances is fair since in practice we are particularly interested in instances that can be realized with few block crossings.
	First, pick $\numchars$, $n$, a probability $p$, and a number $\beta$: we generate random instances that have optimum at most $\beta$ as follows.
	Start from the identity permutation and sample $\beta$ uniformly-random block crossings:
	this results in a sequence of $\beta+1$ permutations.
	Now generate $n$ meetings: pick, for each one independently, one of the permutations at random and then $c$ adjacent characters from this permutation at random, where $c$ is binomially distributed with success probability $p$ so as to match the uniform model; put these meetings in the order of the permutations they come from.
	By construction, these instances have a solution with (at most) $\beta$ block crossings.

\begin{figure}[t]
\centering
\includegraphics{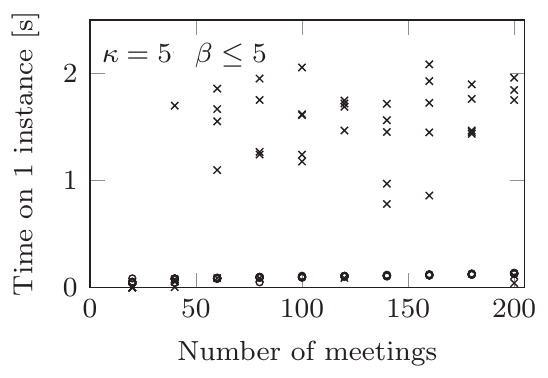}
\hfill
\includegraphics{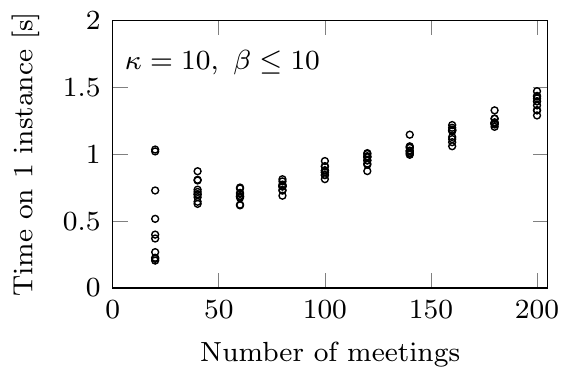}
\caption{Circular marks: runtime of \aSAT{} on random instances from the small-opt model with $p=0.5$, and $\numchars=5$ (left) and $\numchars=10$ (right).
Crosses, left plot: runtime of \aID{}.
It is highly variable and practically dominated by \aSAT{}.
\label{fig:sat_runtime_small}}
\end{figure}

Figure~\ref{fig:sat_runtime_small} shows that \aSAT{} and \aID{} can solve much larger instances of this kind.
This is as expected, since $\beta$ directly bounds the branching depth of \aID{} and the number of permutations required by \aSAT{}.
We see that \aSAT{} practically dominates \aID{}; the only reason to use \aID{} is if no high-quality SAT solver is available, or if memory usage is important (the redeeming quality of \aID{}).

\paragraph{Memory Usage.}

Table~\ref{tab:mem} shows that the (peak) memory usage of the algorithms is quite different.
\aID{} is implemented to branch with a single data structure.
This is good for runtime (no copying), and has the additional benefit that memory usage is very low.
In fact, it is hardly impacted by recursion depth since we use iterative deepening depth-first search and, rather than having the entire data structure at each level of the recursion, there is only a small stack frame.

The memory usage of \aSAT{} increases significantly with $\lambda$ and the overhead is more than for \aID{}: this is because of the large, explicitly-stated SAT formulas and the use of a general-purpose SAT solver.

For small $\numchars$, \aDP{} uses less memory than \aSAT{} due to the latter's overhead.
However, \aDP{} clearly uses the most memory as $\numchars$ increases, since it quite fundamentally relies on the memoization of a large recurrence.
Its memory usage in fact limits the number of characters that can be
supported~-- in practice by the available memory, but even more
generally by the memory architecture of normal environments.

\medskip

\begin{table}[t]
\begin{center}
\begin{tabular}{cScS}
\toprule
$\numchars$ & {\quad\quad \aID{}\quad\quad} & {\quad\quad \aSAT{}\quad\quad} & {\quad\quad \aDP{}\quad\quad} \\
\midrule
    5 &  0.69  &  $ 13$ -- $ 31$  & 1.2     \\
    6 &  0.69  &  $ 44$ -- $ 48$  & 1.7     \\
    7 &  0.69  &  $ 64$ -- $ 76$  & 8.3     \\
    8 &  0.69  &  $110$ -- $218$  & 64.0    \\
    9 &  0.70  &  $338$ -- $422$  & 645.9   \\
\midrule
 1000 &  4.00  & \multicolumn{1}{c}{$\times$} & \multicolumn{1}{c}{$\times$} \\
\bottomrule
\end{tabular}
\end{center}
\caption{Memory usage in MB on uniform random instances ($p=0.5$) with 100 meetings and a variable number of characters $\numchars$.
Only \aSAT{}'s memory usage varies considerably over different instances with the same number of characters.
\label{tab:mem}}
\end{table}

\paragraph{Concluding Remarks.}

We conclude with some practical advice about picking an algorithm.
The first consideration is a hard constraint: if concurrent meetings are required, \aID{} and \aDP{} are disqualified and \aSAT{} remains as a fine default.
We now assume concurrent meetings are not required.

If the number of characters is small, the use of \aDP{} is clearly preferred.
This algorithm can truly be considered fixed-parameter tractable in the number of characters $\numchars$.
However, the dependence on $\numchars$ includes factorial space, which makes it impractical to run the algorithm on personal computers beyond $\numchars=10$ and impossible to run at all for even a few characters more than that.
Many real-world instances have too many characters for \aDP{}.

The runtime of both \aID{} and \aSAT{} depends heavily on the number of block crossings in the optimum.
For very small optimum, \aID{} can be faster, but only if the number of characters is also quite small: there is still the branching factor of $\numchars!$.
If memory is a problem and the optimum is small, then \aID{} is an option, but in general \aSAT{} is vastly preferable.

As a final remark, we note that all these implementations are single-threaded, and as such achieve only ``25\%'' utilization on our quad-core test machine.
\aID{} could be trivially parallelized by dividing the search space;
\aDP{} is trickier to parallelize from an engineering perspective.
It would be possible to use a parallelized SAT solver, like PMSat~\cite{gil2008pmsat} or HordeSAT~\cite{balyo2015hordesat}.
However, it is not clear a~priori how effective those would be for our specific SAT formulas.
\section{Conclusion}
\label{sec:conclusion}

In this paper we have presented a SAT-based algorithm for computing block-crossing optimal storyline visualizations and extensive experimentation on random instances.
We have demonstrated that on some real-world instances (in particular, the
movies), \aSAT{} has runtime similar to the ILP of Gronemann et al., who optimize pairwise crossings.
For other instances (the books), \aSAT{} fares poorly.
We have also evaluated implementations of two further algorithms for storyline block crossing optimization.

For future work, it would be interesting to perform further algorithm engineering on \aSAT{}. %
In particular, it may be possible to handle the birth/death of characters more efficiently or to better integrate with SAT algorithms.

In a different direction, %
one might use an ILP solver on a model very similar to that of Section~\ref{sec:sat}.
This would, for example, enable us to minimize the number of pairwise crossings subject to the number of block crossings being optimal.
However, preliminary experiments showed very poor performance.

From a graphic design perspective, optimizing for block crossings
intuitively makes sense.  However, we are not aware of any user
studies that investigated whether block crossings are good, and what
the trade-offs are.
For example, is a $4 \times 4$ block crossing equally bad as a $2 \times 8$ block crossing?

\paragraph{Acknowledgments.}
We thank Martin Gronemann for providing the exact input files used in
the experiments of~\cite{Gronemann2016}.

\bibliographystyle{splncs03}
\bibliography{abbrv,slbc}

\newpage
\appendix
\section*{\appendixname}

\section{Complete SAT formulation}
\label{app:sat-complete}

In the following, we describe our complete SAT formula in conjunctive normal form.
We do not justify the effect of the clauses here, as this was already done in
Section~\ref{sec:sat}.
To be able to link the clauses with their description, the title of the
paragraphs and order of the clauses are the same as in the main text.

Remember that the input consists of a storyline $\CS=(C,M,E)$ and the number of
permutations $\lambda$.
The number $\mu$ of meeting groups and their composition (represented by
$\CM_\ell$ and $\CL_i$), which is used in the definition of the clauses, can be
derived from $\CS$ easily.

\bgroup
\def\arraystretch{1.2}
\newcommand{\addverticalspace}[1]{%
    \raisebox{0pt}[\dimexpr \height+0.1cm\relax][\dimexpr \depth+0.1cm\relax]{$#1$}%
}
\paragraph{Variables.}
\[
\]

\paragraph{Describing the Permutations.}
\[%
\]

\paragraph{Crossings Between Characters.}
\[%
\]

\paragraph{Block Crossings.}
\[%
\]

\paragraph{Meeting Groups.}
\[%
\]

\egroup

\section{Example Drawings}
\label{app:drawings}

On the following pages, we show our solutions for the movie instances described
in the paper.
For some of the movies, the storyline is split into multiple rows or onto multiple pages.
In this case the parts from top to bottom of a page need to be recombined from left to right.

\begin{sidewaysfigure}[ht]
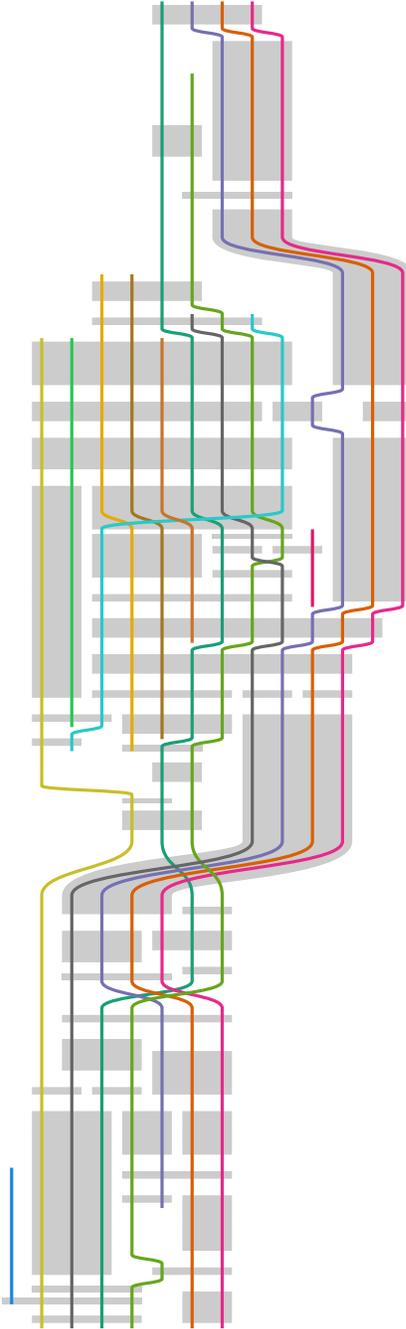

    \centering
    \tikzset{external/export=true}
    %
    \tikzset{external/export=false}
    \caption{A block-crossing optimal drawing of The Matrix.
    \label{fig:thematrix-complete}}
\end{sidewaysfigure}

\begin{sidewaysfigure}[ht]
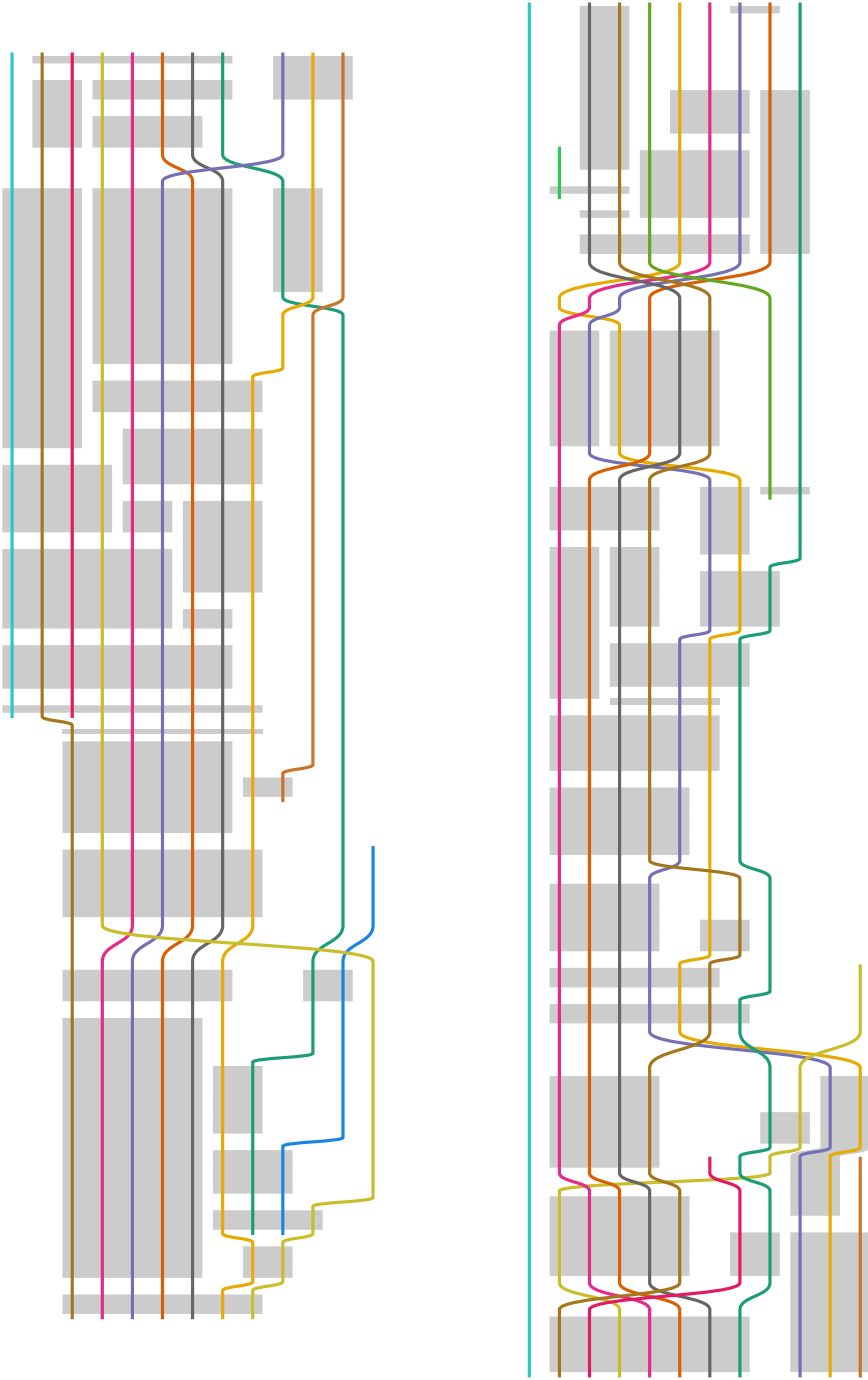

    \centering
    \tikzset{external/export=true}
    %
    \tikzset{external/export=false}
    \caption{A block-crossing optimal drawing of Star Wars.
    \label{fig:star-wars}}
\end{sidewaysfigure}

\begin{sidewaysfigure}[ht]
    \centering
    \tikzset{external/export=true}
    %
    \tikzset{external/export=false}
    \caption{A block-crossing optimal drawing of Inception; part 1 of 2.
    \label{fig:inception-1}}
\end{sidewaysfigure}

\begin{sidewaysfigure}[ht]
    \centering
    \tikzset{external/export=true}
    %
    \tikzset{external/export=false}
    \caption{A block-crossing optimal drawing of Inception; part 2 of 2.
    \label{fig:inception-2}}
\end{sidewaysfigure}

\end{document}